\documentclass[times,twocolumn, 10pt]{article} 
\usepackage{amssymb}     
\usepackage{latexsym}    
\usepackage{amsfonts} 
\usepackage{epsfig} 

\newtheorem{Th}{Theorem}[section]
\newtheorem{theorem}[Th]{Theorem} 
\newtheorem{proposition}[Th]{Proposition}     
\newtheorem{lemma}[Th]{Lemma}
\newtheorem{corollary}[Th]{Corollary}
\newtheorem{definition}[Th]{Definition}

\newcommand{\bpf}{\noindent {\bf Proof:} }

\def\endproof{\hfill$\Box$} 

\newcommand{\lto}{\looparrowright}
\newcommand{\rto}{\looparrowleft}

\newcommand{\bit}{\begin{itemize}}
\newcommand{\eit}{\end{itemize}\par\noindent}
\newcommand{\ben}{\begin{enumerate}}
\newcommand{\een}{\end{enumerate}\par\noindent}  
\newcommand{\beq}{\begin{equation}}
\newcommand{\eeq}{\end{equation}\par\noindent}
\newcommand{\beqa}{\begin{eqnarray*}}
\newcommand{\eeqa}{\end{eqnarray*}\par\noindent}  
\newcommand{\beqn}{\begin{eqnarray}}  
\newcommand{\eeqn}{\end{eqnarray}\par\noindent}          
\newcommand{\bla}{\left\{\begin{array}{l}}
\newcommand{\ela}{\end{array}\right.}       
\newcommand{\bra}{\left\{\begin{array}{r}} 
\newcommand{\era}{\end{array}\right.}

\title{{The Logic of Entanglement}} 
\author{Bob Coecke\\ 
{\small Oxford University Computing Laboratory,}\\
{\small Wolfson Building, Parks Road, OX1 3QD Oxford, UK.}\\
{\small\texttt{coecke@comlab.ox.ac.uk}}}
\date{}
\begin{document} 
\maketitle

\noindent
{\footnotesize
{\bf Abstract.}~~We expose the information flow capabilities of pure
bipartite entanglement as a theorem --- which embodies the exact
statement on the `seemingly acausal flow of information'  in protocols
such as teleportation \cite{Schumacher}. We use this theorem to re-design and
analyze known protocols (e.g.~logic gate
teleportation \cite{Gottesman} and entanglement swapping \cite{Swap}) and show how to
produce some new ones (e.g.~parallel composition of logic gates). We also show how
our results extend to the multipartite case and how they indicate that entanglement
can be measured in terms of `information flow capabilities'.   Ultimately, we propose
a scheme for automated design of protocols involving measurements, local unitary 
transformations and classical communication. 

}

\section{Introduction}

Entanglement has always been a primal ingredient of
fundamental research in quantum theory, and more recently, quantum computing. 
By studying it we aim at understanding the operational/physical 
significance of the use of the Hilbert space tensor product for the description of
compound quantum systems. Many typical quantum phenomena are indeed due to
compound quantum systems being described within the tensor product
${\cal H}_1\otimes{\cal H}_2$ and not within a direct sum ${\cal H}_1\oplus{\cal
H}_2$.

\smallskip
In this paper we reveal a new structural ingredient of the supposedly well-understood
{\it pure bipartite entanglement}, that is, we present a new theorem about the tensor
product of Hilbert spaces.  It 
identifies a `virtual flow of information' in so-called {\it entanglement
specification networks}. For example, it is exactly this flow of information which
embodies {\it teleporting\,}
\cite{BBC} an unknown state from one physical carrier to another. 
Furthermore, our theorem (nontrivially)
extends to {\it multipartite entanglement}. We also argue that it provides a new
way of conceiving entanglement itself and hence of {\it measuring
entanglement}:
\begin{center}
entanglement $\,\equiv\,$ information flow capabilities   
\end{center} 
Indeed, our result enables {\it reasoning about quantum information flow\,} without explicitly considering classical
information flow --- this despite the impossibility of transmitting quantum information through entanglement without the
use of a classical channel.   
 
\smallskip
Using our theorem we can evidently reconstruct protocols such as {\it logic
gate teleportation\,}
\cite{Gottesman} and {\it entanglement swapping\,} \cite{Swap}.  It moreover allows
smooth generation of new protocols, of which we provide an example, namely the
conversion of accumulation of inaccuracies causing `sequential composition' into
{\it fault-tolerant\,} `parallel composition' \cite{Preskill}. 
Indeed, when combing our new insights on the flow of information through entanglement
with a model for the flow of classical information
we obtain a powerful tool for designing protocols involving entanglement.  

\smallskip
An extended version of this paper is available as a
research report \cite{RR}. It contains details of
proofs, other/larger pictures, other references, other applications and
some indications of connections with logic, proof theory
and functional programming. 

\section{Classical information flow}\label{sec:class}

\smallskip
By the spectral theorem any non-degenerated measurement on a quantum system described in a $n$-dimensional
complex Hilbert space ${\cal H}$ has the shape
\[
M=x_1\cdot {\rm P}_1+\ldots+x_n\cdot {\rm P}_n\,.
\]
Since the values $x_1,\ldots,x_n$ can be conceived as merely being tokens distinguishing 
the projectors ${\rm P}_1,\ldots,{\rm P}_n$ in the
above sum we can abstract over them and conceive such
a measurement as a set
\[
M\simeq\{{\rm P}_1,\ldots,{\rm P}_n\}
\]
of $n$ mutually orthogonal projectors which each project on a
one-dimensional subspace of ${\cal H}$. Hence, by von
Neumann's projection postulate, a measurement can be
conceived as the system being  subjected to an {\it
action\,} ${\rm P}_i$ and the observer being informed
about which action happened  (e.g.~by receiving the token
$x_i$).

\smallskip
In most quantum information 
protocols the indeterminism of measurements necessitates a
flow of classical information e.g.~the 2-bit classical
channel required for teleportation \cite{BBC}.  We want to
separate this {\it classical information flow\,} from what 
we aim to identify as the {\it quantum information flow}.
Consider a protocol involving local unitary operations,
(non-local) measurements and classical
communication e.g.~teleportation: 

\vspace{2.0mm}\noindent{\footnotesize 
\begin{minipage}[b]{1\linewidth}  
\centering{\epsfig{figure=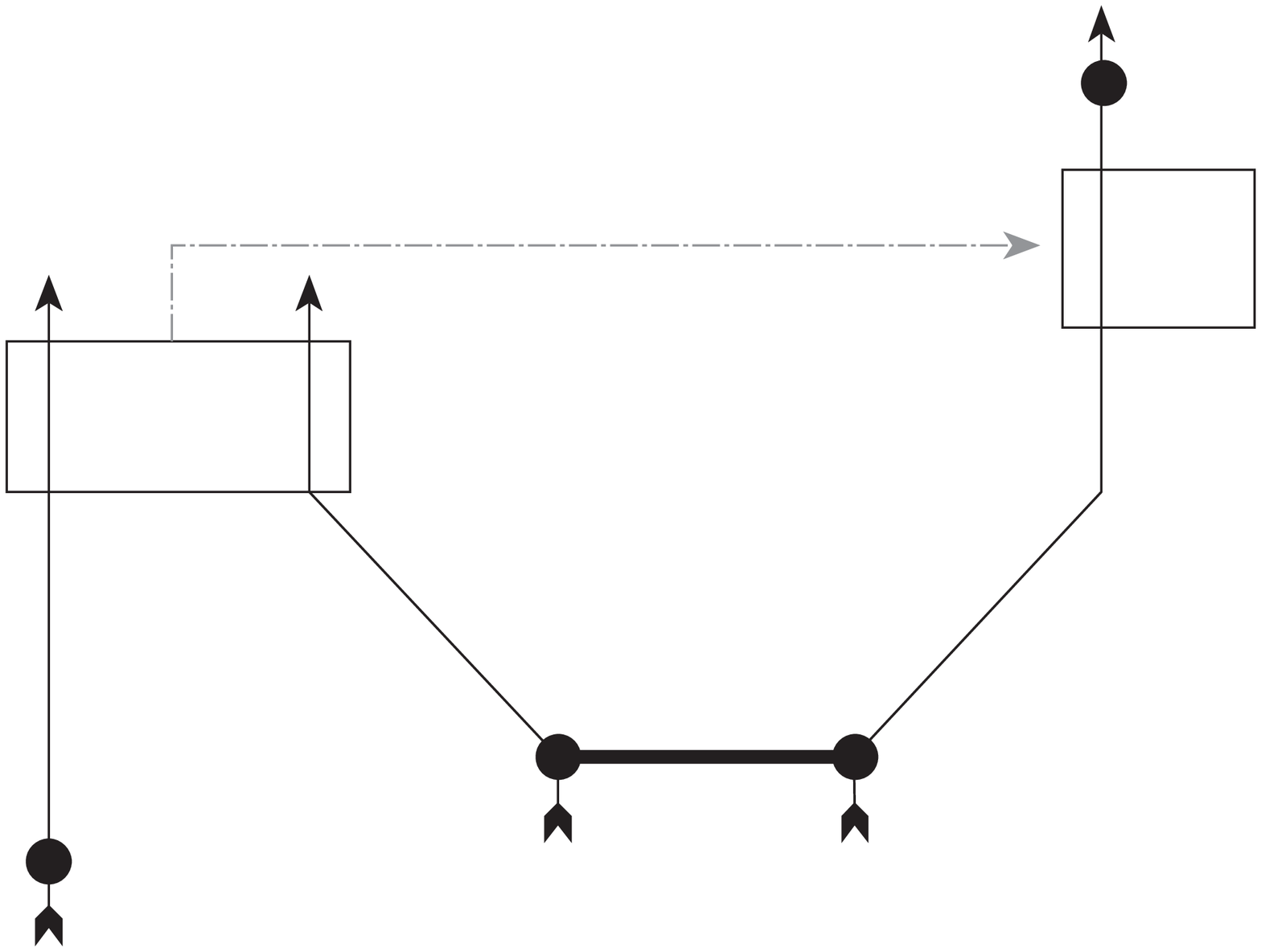,width=110pt}}     
  
\begin{picture}(110,0)   
\put(51,30){$\Psi_{\!E\!P\!R}$}  
\put(5.0,54){$M_{\!Bell}$}  
\put(98.1,69){$U_{\!x\!z}$}  
\put(55,74){${\!x\!z}$}  
\put(101,87){$\phi$}  
\put(-5,13){$\phi$} 
\end{picture}  
\end{minipage}}

\vspace{-3.5mm}\noindent
We can decompose such a protocol in 
\ben
\item a {\it tree\,} with the consecutive operations as nodes,
and, in case of a measurement, the emerging branches being
labeled by tokens representing the projectors;
\item  the configuration of the operations in terms of
the time when they are applied and the subsystem
to which they apply. 
\een
Hence we abstract over spatial dynamics.
The nodes in the tree are connected to the boxes in the {\it configuration
picture\,} by their temporal coincidence. 
For teleportation we thus obtain
 
\vspace{1.0mm}\noindent{\footnotesize  
\begin{minipage}[b]{1\linewidth}  
\centering{\epsfig{figure=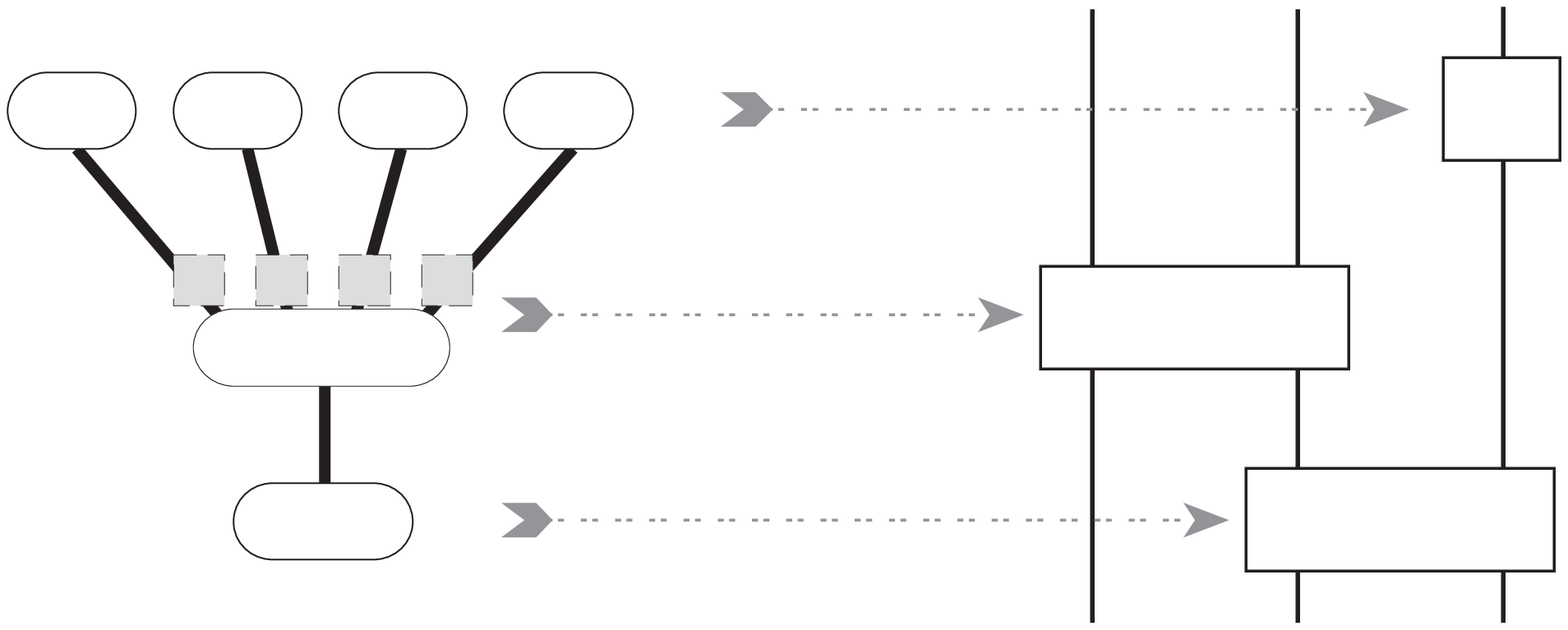,width=205pt}}      
  
\begin{picture}(205,0)   
\put(32,21.5){$\Psi_{\!E\!P\!R}$}     
\put(31,44.8){$M_{\!Bell}$}  
\put(3,75.8){$U_{\!00}$}   
\put(25,75.8){$U_{\!01}$}    
\put(47,75.8){$U_{\!10}$}   
\put(69,75.8){$U_{\!11}$}   
\put(22.8,55.2){\scriptsize${}_{0\hspace{-0.5pt}0}$}    
\put(33.6,55.2){\scriptsize${}_{0\hspace{-0.5pt}1}$}      
\put(44.2,55.2){\scriptsize${}_{1\hspace{-0.5pt}0}$}    
\put(55.55,55.2){\scriptsize${}_{1\hspace{-0.7pt}1}$} 
\put(194.5,75.8){$...$}   
\put(153.2,48.8){$...$}  
\put(180.7,21.5){$...$}     
\end{picture}  
\end{minipage}}

\vspace{-3mm}\noindent
Classical communication is encoded in the tree as the dependency
of operations on the labels on the branches below it e.g.~the
dependency of the operation $U_{xz}$ on the variable $xz$ stands for the 2-bit
classical channel required for teleportation. We will also
replace any initial state $\Psi$ by the projector
${\rm P}_\Psi$ on it, which can be conceived as its {\it
preparation\,} e.g.~${\rm P}_{\!E\!P\!R}$ is the 
preparation of an EPR-pair.  It should be clear that for each path from the
root of the tree to a leaf, by `filling in the
operations on the included nodes in the corresponding boxes of the
configuration picture', we obtain a network involving only local unitary
operations and (non-local) projectors  e.g. one network
 
\vspace{1.3mm}\noindent{\footnotesize  
\begin{minipage}[b]{1\linewidth}  
\centering{\epsfig{figure=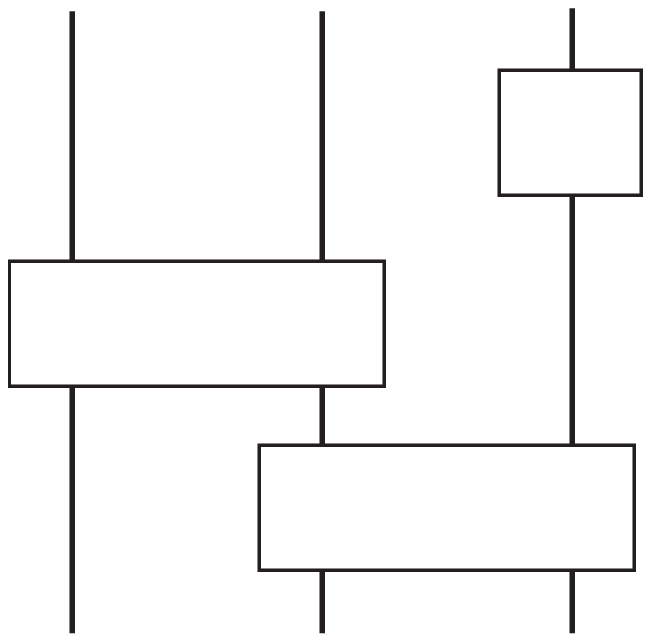,width=70pt}}      
  
\begin{picture}(70,0)   
\put(40,21.5){${\rm P}_{\!E\!P\!R}$}      
\put(16,42.5){${\rm P}_{\!xz}$}   
\put(56.8,63.5){$U_{\!xz}$}    
\end{picture}  
\end{minipage}}

\vspace{-3mm}\noindent
for each of the four values $xz$ takes.
It will be these networks (from which we extracted the classical information
flow) for which we will reveal the quantum information flow. Hence each projector in
it which is not a preparation is to be conceived {\it conditionally}. 


\section{Bipartite entanglement} 

Let ${\cal H}_1$ and ${\cal H}_2$ be two finite
dimensional complex Hilbert spaces. The elements of ${\cal
H}_1\otimes{\cal H}_2$ are in bijective correspondence with those
of ${\cal H}_1\!\to\!{\cal H}_2$, the vector space of {\it linear
maps\,} with domain
${\cal H}_1$ and codomain ${\cal H}_2$, and also with those of ${\cal
H}_1\!\lto\!{\cal H}_2$, the vector space  of {\it anti-linear maps\,}
with domain ${\cal H}_1$ and codomain ${\cal H}_2$. Given a base 
$\{{e}_{\alpha}^{(1)}\}_{\alpha}$ of ${\cal H}_1$ and a base $\{{e}_{\beta}^{(2)}\}_{\beta}$ of ${\cal H}_2$ 
this can easily be seen through the correspondences
\beqa 
\sum_{\alpha\beta}m_{\alpha\beta}\,
\langle{e}_{\alpha}^{(1)}\mid-\rangle\cdot
{e}_{\beta}^{(2)}\!\!\!&\stackrel{\rm L}{\simeq}&\!\!
\sum_{\alpha\beta}m_{\alpha\beta}\cdot {e}_{\alpha}^{(1)}\otimes{e}_{\beta}^{(2)}\\
\sum_{\alpha\beta}m_{\alpha\beta}\,
\langle-\mid{e}_{\alpha}^{(1)}\rangle\cdot {e}_{\beta}^{(2)}
\!\!\!&\stackrel{\rm aL}{\simeq}&\!\!
\sum_{\alpha\beta}m_{\alpha\beta}\cdot {e}_{\alpha}^{(1)}\otimes{e}_{\beta}^{(2)}
\eeqa
where $(m_{\alpha\beta})_{\alpha\beta}$ is the {\it matrix\,} of the  
corresponding function in bases $\{{e}_{\alpha}^{(1)}\}_{\alpha}$ and
$\{{e}_{\alpha}^{(2)}\}_{\beta}$ and where by 
\[
\langle{e}_{\alpha}^{(1)}\mid-\rangle:{\cal H}_1\to{\cal H}_2
\quad{\rm and}\quad
\langle-\mid{e}_{\alpha}^{(1)}\rangle:{\cal H}_1\lto{\cal H}_2
\]
we denote the functionals which are respectively the linear and the anti-linear duals
to the vector
${e}_{\alpha}^{(1)}$.  While the second correspondence does not depend on the choice
of
$\{{e}_{\alpha}^{(1)}\}_{\alpha}$ the first one does since
\[
\langle c\cdot{e}_{\alpha}^{(1)}\!\!\mid-\rangle=\bar{c}\cdot\langle
{e}_{\alpha}^{(1)}\!\!\mid-\rangle
\ \ {\rm and}\ \
\langle -\!\!\mid c\cdot{e}_{\alpha}^{(1)}\rangle={c}\cdot\langle
-\!\!\mid {e}_{\alpha}^{(1)}\rangle.
\]
We can now represent the {\it states\,} of 
${\cal H}_1\otimes{\cal H}_2$ by functions in 
${\cal H}_1\!\to\!{\cal H}_2$ or in ${\cal H}_1\!\lto\!{\cal H}_2$, and vice
versa, these functions represent states of ${\cal
H}_1\otimes{\cal H}_2$. Omitting normalization constants,  
an attitude we will abide by throughout this paper,
examples of linear maps encoding states are: 
\beqa
{\rm id}:=\left(\begin{array}{cc}
1&0\\
0&1
\end{array}\right)\ &\stackrel{\rm L}{\simeq}& \  
|00\rangle+|11\rangle\\
\pi:=\left(\begin{array}{rr}
0&1\\
1&0
\end{array}\right)\ &\stackrel{\rm L}{\simeq}& \ 
|01\rangle+|10\rangle\\
{\rm id}^*:=\left(\begin{array}{rr}
1&0\\
0&\!\!\!\!\!-\!1
\end{array}\right)\ &\stackrel{\rm L}{\simeq}& \ 
|00\rangle-|11\rangle\\
\pi^*:=\left(\begin{array}{rr}
0&\!\!\!\!\!-\!1\\
1&0
\end{array}\right)\ &\stackrel{\rm L}{\simeq}& \ 
|01\rangle-|10\rangle
\eeqa
These four functions which encode the {\it Bell-base}
states are almost the {\it Pauli matrices}
\[
\sigma_x\equiv X:=\pi
\quad\quad
\sigma_y\equiv Y:=i\pi^*
\quad\quad
\sigma_z\equiv Z:={\rm id}^*  
\]
plus the {\it identity\,} which itself encodes the {\it EPR-state}. 
We can also encode
each projector 
\[ 
{\rm P}_\Psi:{\cal H}_1\otimes{\cal H}_2\to{\cal H}_1\otimes{\cal
H}_2::\Phi\mapsto
\langle\Psi\mid\Phi\rangle\cdot\Psi
\]
with $\Psi\in{\cal H}_1\otimes{\cal H}_2$ by a function
either in ${\cal H}_1\!\to\!{\cal H}_2$ or ${\cal H}_1\!\lto\!{\cal H}_2$.
Hence we can use these (linear or anti-linear) {\it functional labels\,} 
both to
denotate the states of ${\cal H}_1\otimes{\cal H}_2$ and the
projectors on elements of ${\cal H}_1\otimes{\cal H}_2$. 
We introduce a graphical notation which incarnates this.

\vspace{2.5mm}\noindent
\begin{minipage}[b]{1\linewidth}  
\centering{\epsfig{figure=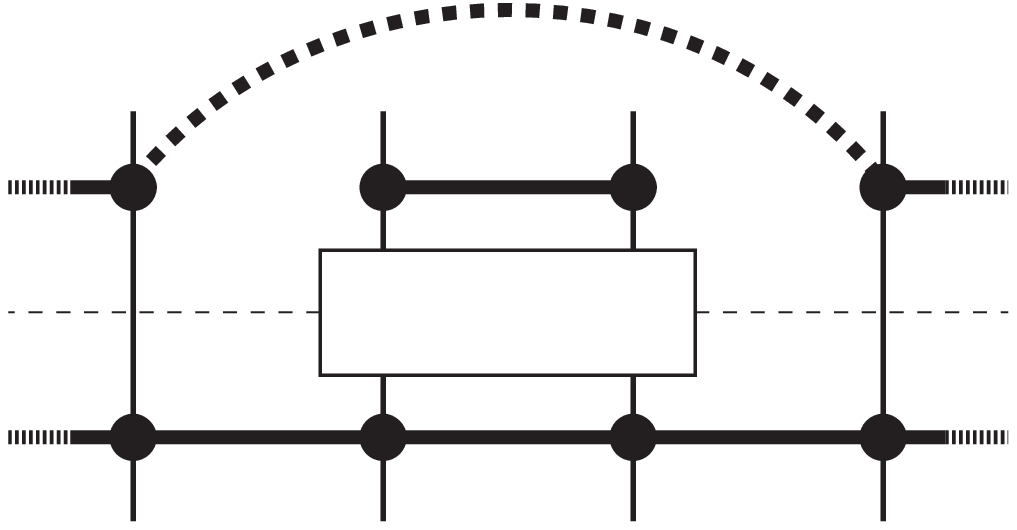,width=100pt}}  

\begin{picture}(100,0)     
\put(48,30.5){$f$}
\put(48,49.5){$f$}
\put(34,4){\small${\cal H}_1$} 
\put(59,4){\small${\cal H}_2$}
\put(-15,15){\vector(0,1){40}}
\put(-39,52){$time$}
\end{picture}
\end{minipage}

\vspace{-1mm}\noindent  
The box
\raise-2.6pt\hbox{\vspace{2mm}\epsfig{figure=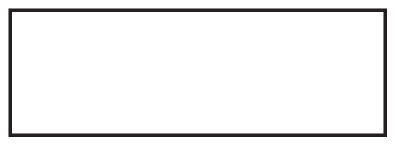,width=28pt}}\hspace{-5.5mm}{\footnotesize$f$}\hspace{4.9mm}
depicts the projector which projects on the bipartite state labeled by
the
\mbox{(anti-)}linear function $f$ and the barbell 
{\,}\raise-3.0pt\hbox{\vspace{2mm}\epsfig{figure=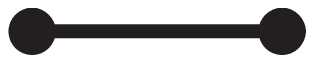,width=25pt}}\hspace{-5mm}\raise2.6pt\hbox{\footnotesize$f$}\hspace{3.4mm}{\,}
depicts that state itself.  Hence the projector 
\raise-2.6pt\hbox{\vspace{2mm}\epsfig{figure=eProjector.eps,width=28pt}}\hspace{-5.5mm}{\footnotesize$f$}\hspace{4.9mm}
acts on the multipartite state represented by 
\raise-1.5pt\hbox{\vspace{2mm}\epsfig{figure=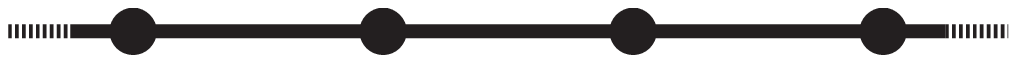,width=100pt}} 
and produces a pure tensor consisting  of (up to a normalization constant) 
{\,}\raise-3.0pt\hbox{\vspace{2mm}\epsfig{figure=eState.eps,width=25pt}}\hspace{-5mm}\raise2.6pt\hbox{\footnotesize$f$}\hspace{3.4mm}{\,}
and some remainder. Hence this picture portrays `preparation of 
the $f$-labeled state'. 

\smallskip
By an {\it entanglement specification network\,} we mean a collection of bipartite projectors
\raise-2.6pt\hbox{\vspace{2mm}\epsfig{figure=eProjector.eps,width=28pt}}\hspace{-5.5mm}{\footnotesize$f$}\hspace{4.9mm}
`configured in space and time' e.g.
  
\vspace{2mm}\noindent
\begin{minipage}[b]{1\linewidth}   
\centering{\epsfig{figure=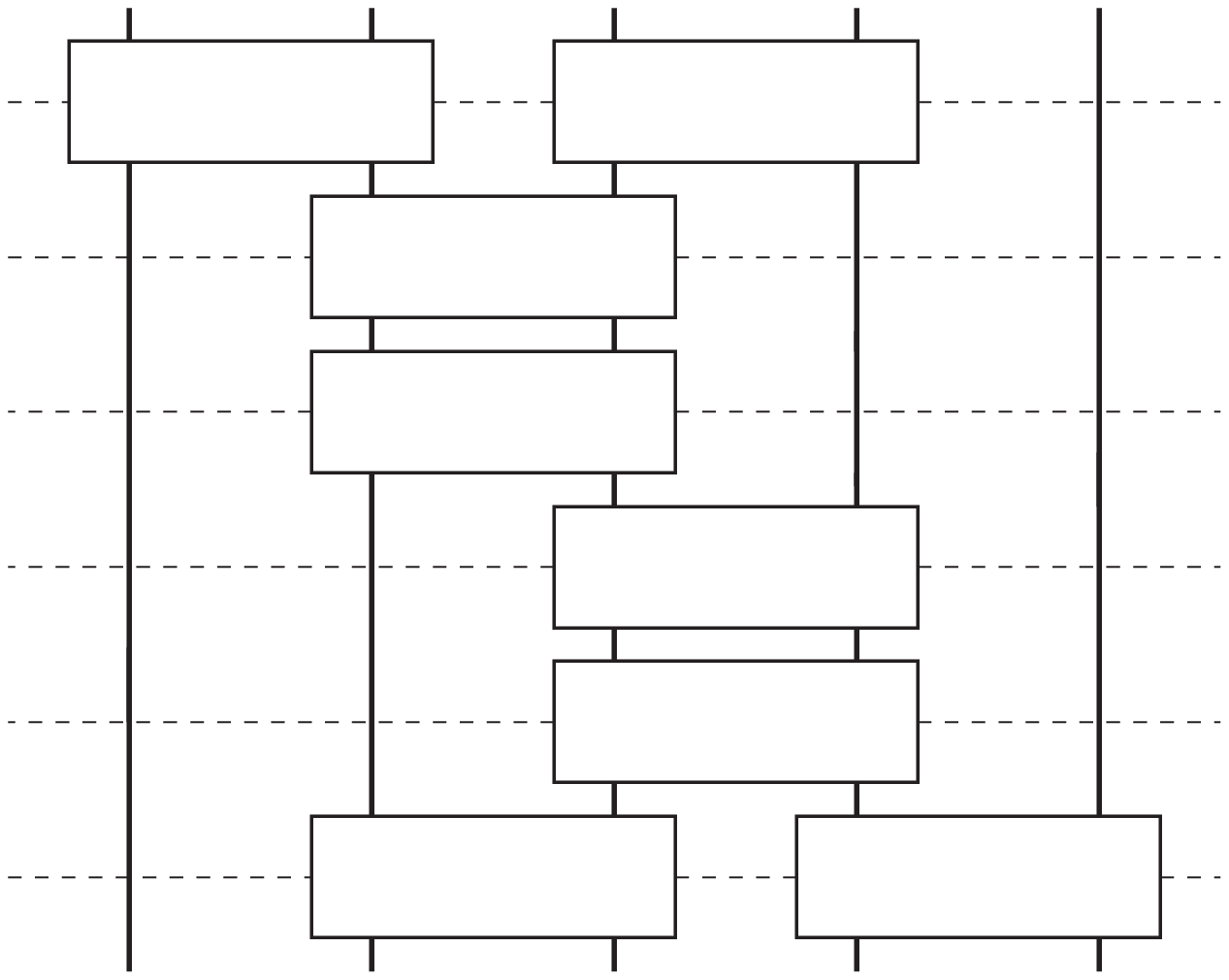,width=125pt}}   

\begin{picture}(125,0)   
\put(-10,20){$1$}
\put(-10,36){$2$}
\put(-10,52){$3$}
\put(-10,68){$4$}
\put(-10,84){$5$}
\put(-10,100){$6$} 
\put(9,100){$-\!\!\!\!-{\scriptstyle f_1}\!\to$}
\put(59,100){$-\!\!\!\!-{\scriptstyle f_3}\!\to$}
\put(34.5,84){$-\!\!\!\!-{\scriptstyle f_2}\!\to$}
\put(34.5,68){$\leftarrow\!{\scriptstyle f_5}-\!\!\!\!-$}
\put(59,52){$\leftarrow\!{\scriptstyle f_4}-\!\!\!\!-$}
\put(59,36){$-\!\!\!\!-{\scriptstyle f_7}\!\to$}
\put(34.5,20){$-\!\!\!\!-{\scriptstyle f_6}\!\to$}
\put(84.5,20){$-\!\!\!\!-{\scriptstyle f_8}\!\to$}
\put(9,2){\small${\cal H}_1$}   
\put(34,2){\small${\cal H}_2$}
\put(59,2){\small${\cal H}_3$}
\put(84,2){\small${\cal H}_4$}
\put(109,2){\small${\cal H}_5$}
\end{picture} 
\end{minipage}   

\vspace{0mm}\noindent
The arrows indicate which of the two Hilbert spaces in ${\cal
H}_1\otimes{\cal H}_j$ is the domain and which is the codomain of the
labeling function. Such a network can also contain local
unitary operations --- which we will represent by a grey square box 
\raise-2.4pt\hbox{\vspace{2mm}\epsfig{figure=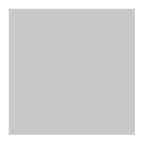,width=10pt}}\hspace{-2.8mm}{\small$U$}\hspace{0.6mm}.
We will refer to the lines labeled by some Hilbert space ${\cal H}_i$
($\simeq$\,time-lines) as {\it
tracks}.
\begin{definition}\em
A {\it path\,} is a line which progresses along the 
tracks either
forward or backward with respect to the actual physical time, and, 
which: (i) 
respects the  four possibilities   

\vspace{-0.5mm}\noindent
\begin{minipage}[b]{1\linewidth}   
\centering{\epsfig{figure=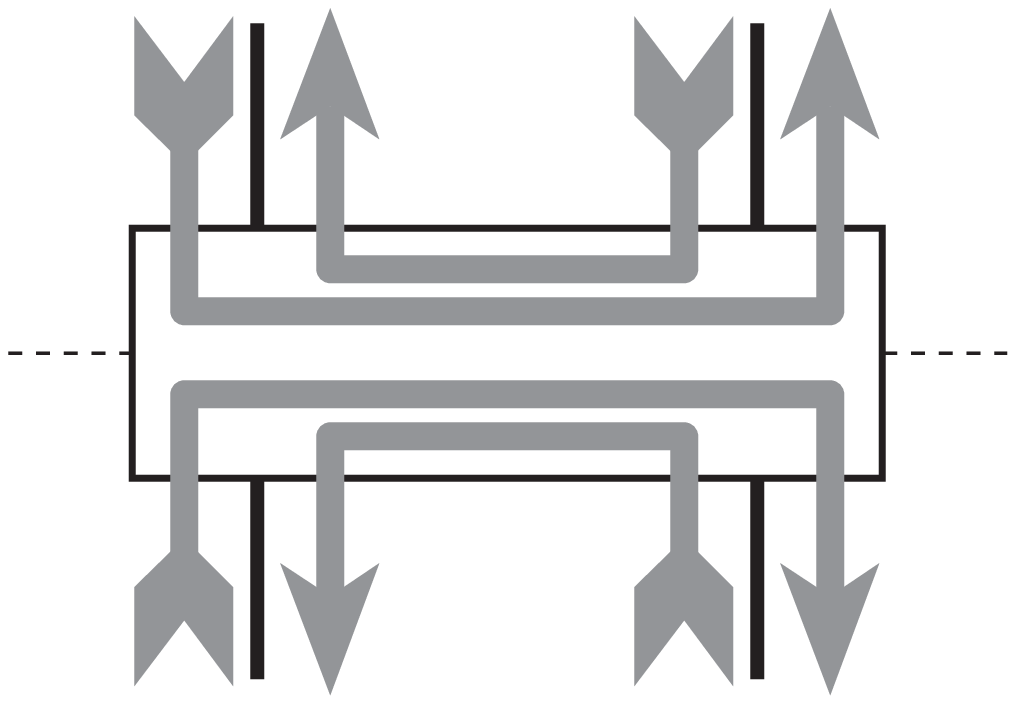,width=60pt}}  
\end{minipage}

\noindent
for entering and leaving a bipartite projector;
(ii) passes
local unitary operations unaltered, that is

\vspace{1.5mm}\noindent
\begin{minipage}[b]{1\linewidth}   
\centering{\epsfig{figure=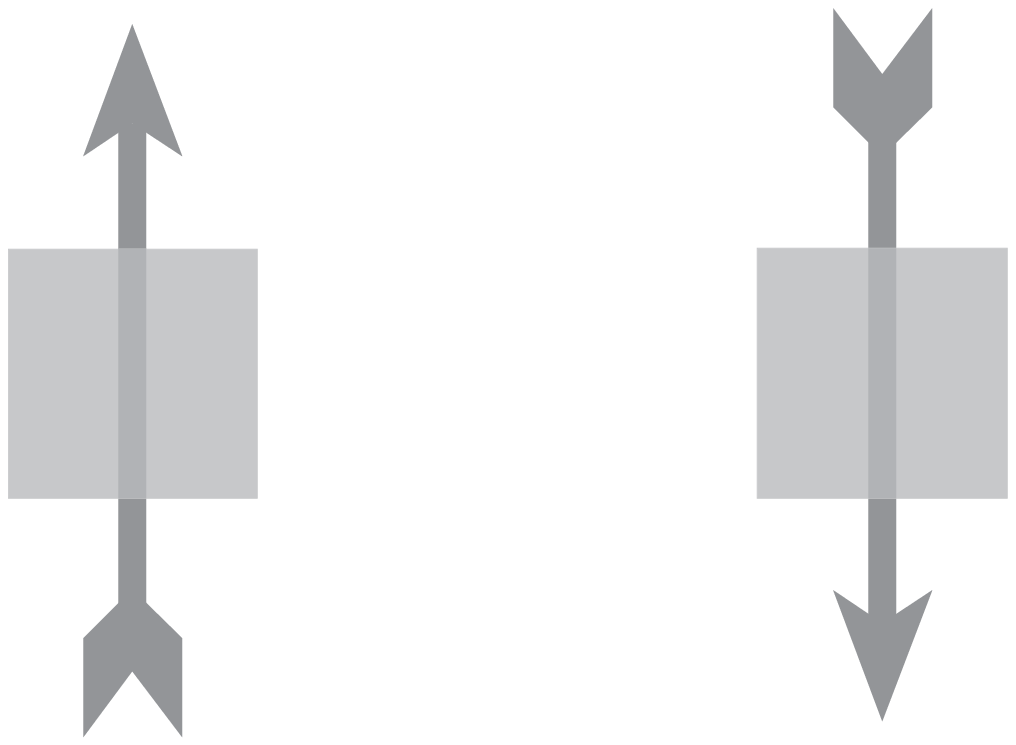,width=60pt}}   
\end{minipage}

\vspace{-0.5mm}\noindent 
in pictures;
(iii) does not end at a time before
any other time which it covers.
\end{definition}
An example of a path is the grey line below. 
 
\vspace{2mm}\noindent
\begin{minipage}[b]{1\linewidth}  
\centering{\epsfig{figure=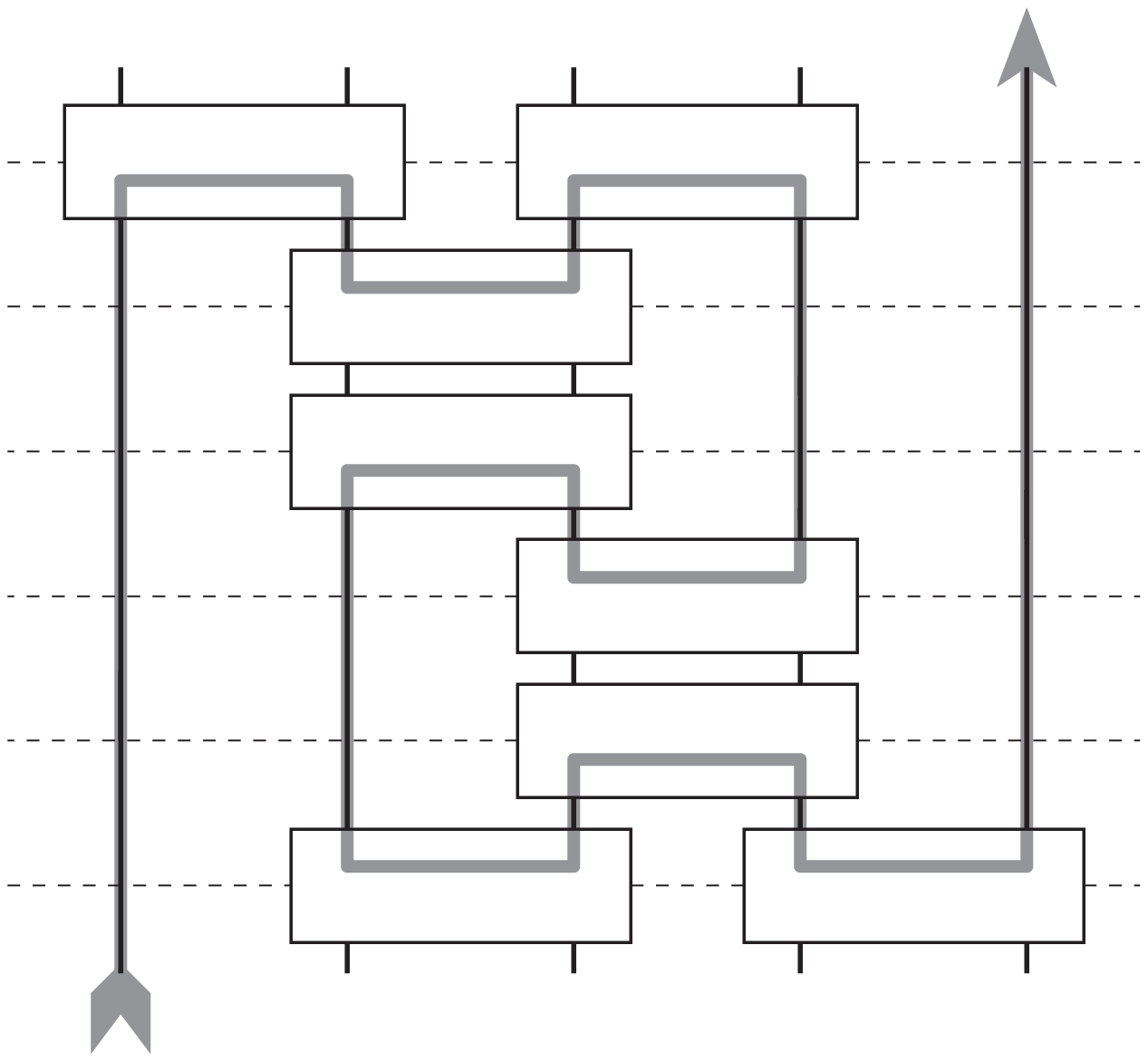,width=125pt}}    

\begin{picture}(125,0)   
\put(-10,28){$1$}
\put(-10,44){$2$}
\put(-10,60){$3$}
\put(-10,76){$4$}
\put(-10,92.5){$5$}
\put(-10,107){$6$} 
\end{picture} 
\end{minipage}   

\vspace{-4mm}\noindent
The notion of a path allows us to make certain predictions
about the output $\Psi_{\bf out}$ of a
network, that is, the state of the whole system after all
projectors have been effectuated.  Before stating  the
theorem we illustrate it on our example.
Let 
\[
\Psi_{\bf in}\,:=\,\phi_{in}\otimes\!\!\!
\sum_{\alpha_2\ldots\alpha_5}\!\!\!
\Phi^{{in}}_{\!\alpha_2\ldots\alpha_5}\!\!\!\cdot
e_{\alpha_2}^{(2)}\!\otimes\! e_{\alpha_3}^{(3)}\!\otimes\! e_{\alpha_4}^{(4)}\!\otimes\!  
e_{\alpha_5}^{(5)}
\]
be its input state.   This input state factors into
the {\it pure factor\,} $\phi_{in}$, which we call the {\it input of
the path}, and a remainder.  

\vspace{-1mm}\noindent
\hspace{-1mm}\begin{minipage}[b]{1\linewidth}  
\centering{\epsfig{figure=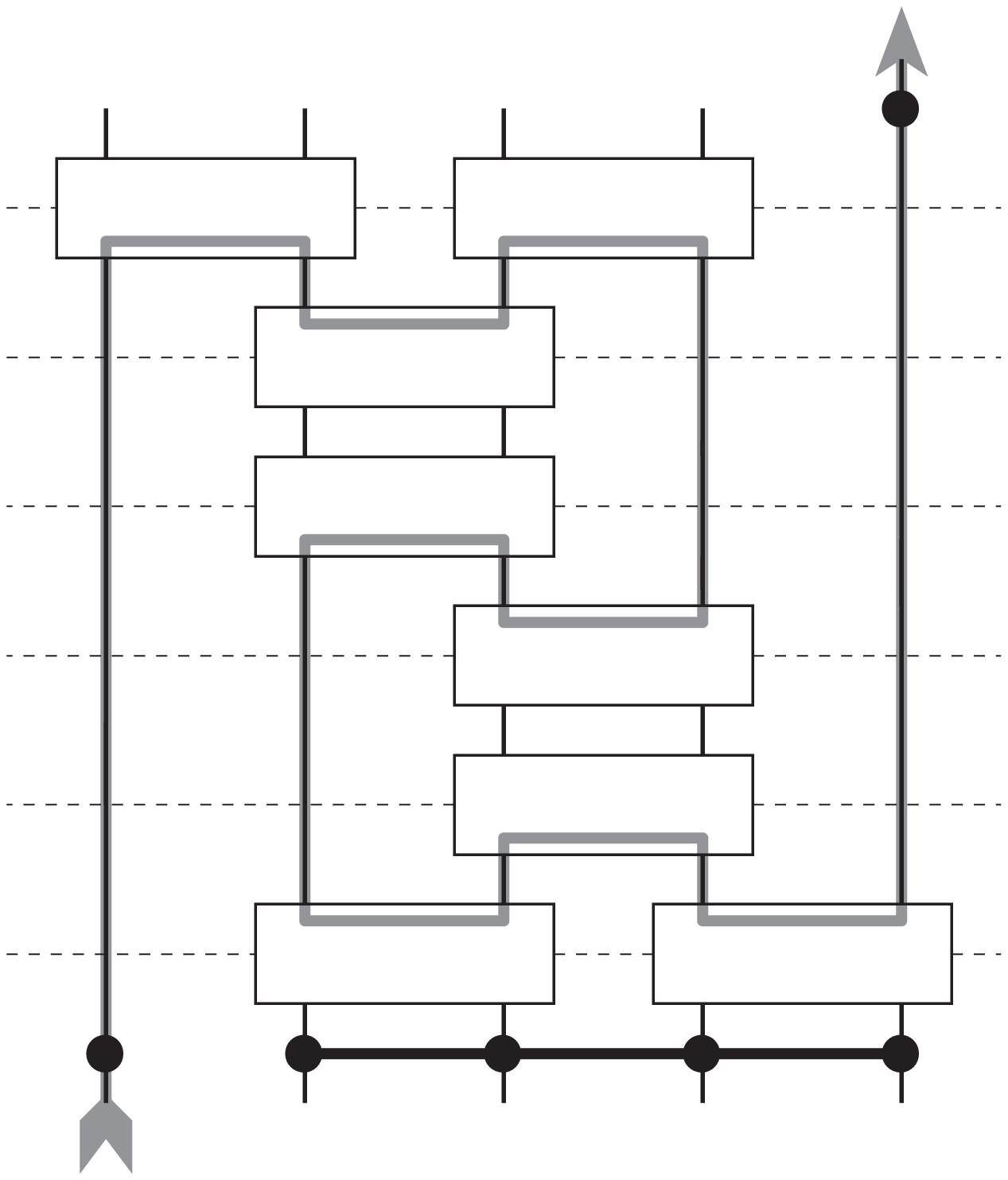,width=133.3pt}}    

\begin{picture}(133.3,0)   
\put(-10,38){$1$}
\put(-10,58){$2$}
\put(-10,78){$3$} 
\put(-10,98){$4$}
\put(-10,118){$5$}
\put(-10,138){$6$}
\put(12,141.5){$-\!\!\!\!-{\scriptstyle f_1}\!\to$}
\put(65,141.5){$-\!\!\!\!-{\scriptstyle f_3}\!\to$}
\put(38.5,118.5){$-\!\!\!\!-{\scriptstyle f_2}\!\to$}
\put(65,78.5){$\leftarrow\!{\scriptstyle f_4}-\!\!\!\!-$}
\put(39,101.5){$\leftarrow\!{\scriptstyle f_5}-\!\!\!\!-$}
\put(39,38.5){$-\!\!\!\!-{\scriptstyle f_6}\!\to$}
\put(65,61.5){$-\!\!\!\!-{\scriptstyle f_7}\!\to$}
\put(91.5,38.5){$-\!\!\!\!-{\scriptstyle f_8}\!\to$} 
\put(35,22){$\underbrace{\mbox{\hspace{3.3cm}}}$} 
%
\put(12,6){\footnotesize$\phi_{in}$}
\put(125.5,152.5){\footnotesize$?[\phi_{out}]$}  
\put(35,5){\scriptsize${\ \ \ \sum
\Phi^{{in}}_{\!\alpha_2\ldots\alpha_5}\!\!\!\!\cdot\!
e_{\alpha_2}^{(2)}\!\otimes\! e_{\alpha_3}^{(3)}\!\otimes\! e_{\alpha_4}^{(4)}\!\otimes\! 
e_{\alpha_5}^{(5)}}$}
\put(30,-1){\scriptsize${\ \ {}_{\alpha_2\ldots\alpha_5}}$} 
\end{picture} 
\vspace{2mm}
\end{minipage}   

\noindent
It should be clear that after
effectuating all projectors we end up with an output which factors in
the bipartite state labeled by
$f_1$, the bipartite state labeled by $f_2$ and a remaining pure factor 
$\phi_{out}$ --- which we call the {\it output of the path}.  Our
theorem (below) predicts that 
\beq\label{eq:1}
\phi_{out}=(f_8\circ f_7\circ f_6\circ f_5\circ f_4\circ 
f_3\circ f_2\circ f_1)(\phi_{in})\,.
\eeq
Be aware of the fact that the functions
$f_1,\ldots,f_8$ are not physical operations but labels obtained via
a purely mathematical isomorphism.
Moreover, the order in which they appear in the composite (\ref{eq:1}) has
no obvious relation to the temporal order of the corresponding
projectors.  Their order in the composite
(\ref{eq:1}) is:
\begin{center}
{\it the order in which the path passes
through them}
\end{center}
--- this despite the fact that the path goes both forward and
backward in physical time. Here's the theorem. 

\begin{lemma}\label{lm:compos}
For $f$, $g$ and $h$ anti-linear maps and $U$ and $V$ unitary
operations we have

\vspace{5mm}\noindent 
\begin{minipage}[b]{1\linewidth}  
\centering{\epsfig{figure=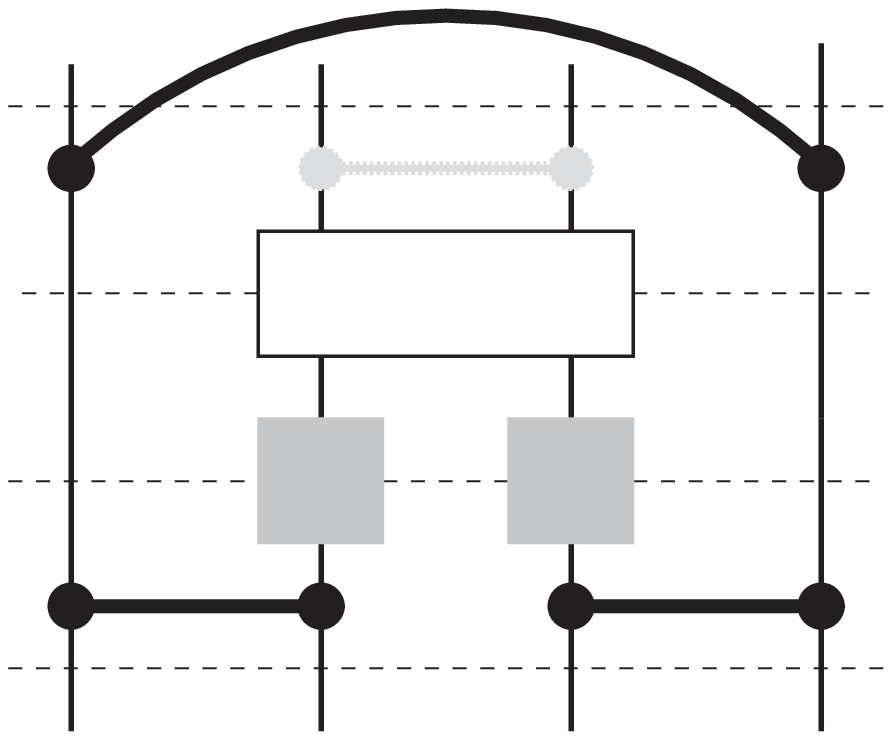,width=165pt}} 

\begin{picture}(165,0)       
\put(81,59){$g$}  
%
\put(44,96.5){$\, h\circ V^\dagger\circ g\circ U\circ f$}    
\put(52,31){$f$}
\put(108,30){$h$} 
\put(66,37){$U$}
\put(93,37){$V$}
\end{picture}
\end{minipage}

\vspace{-6mm}\noindent
\end{lemma}
\bpf Straightforward verification or see \cite{RR} \S 5.1. 
\hfill\endproof
\begin{theorem}\label{compositionalitytheorem}
{\bf (i)} Given are an entanglement specification network and a path.  Assume
that{\rm\,:}
\ben 
\item The order in which the path passes through the projectors is{\,}
${f_1}\Rightarrow {f_2}\Rightarrow\ldots\Rightarrow{f_{k-1}}\Rightarrow{f_k}$.
\item The input of the path is a pure factor $\phi_{in}$\,.
\item $\Psi_{\bf out}$ has a non-zero amplitude{\rm .}
\een
Then the output of the path is (indeed) a pure factor $\phi_{out}$
which is explicitly given by 
\beq\label{eq:2}
\phi_{out}=(f_k\circ f_{k-1}\circ \ldots \circ f_2\circ f_1)(\phi_{in})\,.
\eeq
{\bf (ii)} If the path passes forwardly through 
\raise-2.4pt\hbox{\vspace{2mm}\epsfig{figure=eUnitary.eps,width=10pt}}\hspace{-2.8mm}{\small$U$}\hspace{0.2mm}
then $U$ will be part of the composite {\rm(\ref{eq:2})} and if it passes backwardly
through
\raise-2.4pt\hbox{\vspace{2mm}\epsfig{figure=eUnitary.eps,width=10pt}}\hspace{-2.8mm}{\small$U$}\hspace{0.2mm}
then $U^\dagger$ will be part of the composite {\rm(\ref{eq:2})}.
\end{theorem}
\bpf 
Lemma \ref{lm:compos} is the crucial lemma for the proof. 
For a full proof see \cite{RR} \S 5.  
\hfill\endproof\newline

\noindent  
It might surprise the reader that in the formulation of Theorem 
\ref{compositionalitytheorem} we didn't specify whether 
$f_1,\ldots,f_k$ are either linear or anti-linear, and indeed, we
slightly cheated.  The theorem is only valid for $f_1,\ldots,f_k$
anti-linear. However, in the case that $f_1,\ldots,f_k$ are linear,
in order to make the theorem hold it suffices to conjugate the matrix 
elements of those functional labels for which the path enters 
(and leaves) the corresponding projector `from below'
(see \cite{RR} \S 4.1): 

\vspace{1.5mm}\noindent
\begin{minipage}[b]{1\linewidth}  
\centering{\epsfig{figure=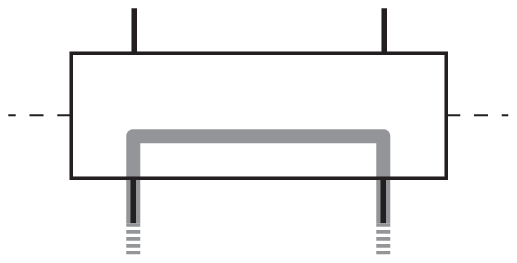,width=52pt}}      
\end{minipage} 

\vspace{-0.5mm}\noindent 
In most practical
examples these matrix elements are real (see below) and hence the above theorem
also holds for linear functional labels.  One also verifies that
if a path passes though a projector in the opposite direction of the
direction of an anti-linear functional label $f$, then we
have to use the {\it adjoint\,} $f^\dagger$ of the anti-linear map $f$ in the 
composite {\rm(\ref{eq:2})} --- the matrix of the adjoint of an anti-linear map
$f^\dagger$ is the transposed of the matrix of $f$ (see \cite{RR}
\S 4.2).  Finally note that we did not specify that at its input a path
should be directed forwardly in physical time, and indeed, the theorem
also holds for paths such as 

\vspace{-0.5mm}\noindent
\begin{minipage}[b]{1\linewidth}  
\centering{\epsfig{figure=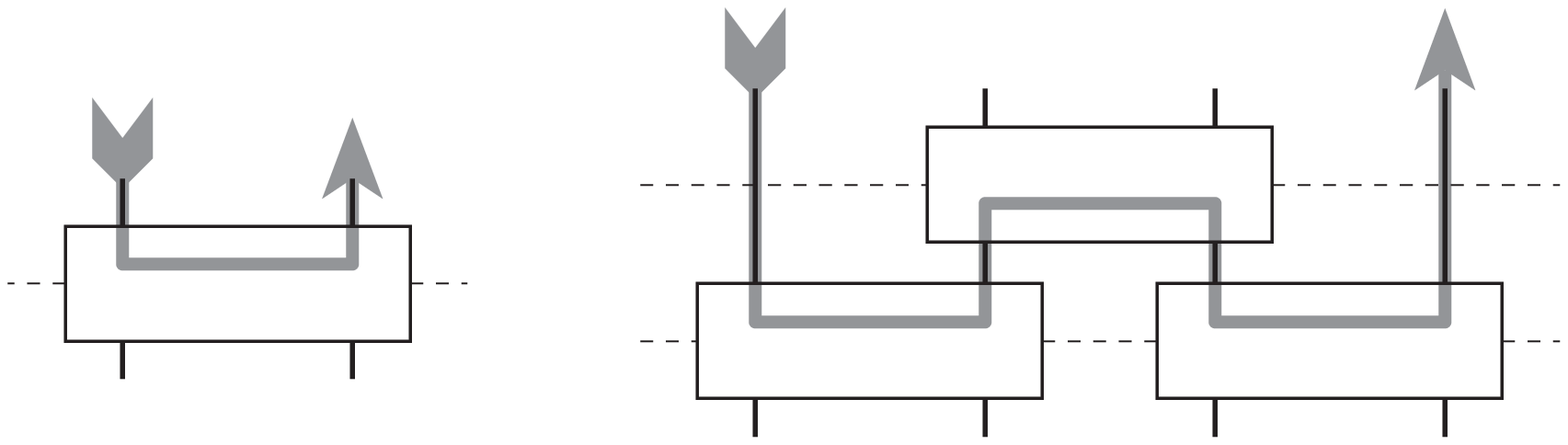,width=165pt}}   
\end{minipage} 

\noindent
We discuss this in Section \ref{sec:outputonly}.

\section{Re-designing teleportation}\label{sec:teleport}

By Theorem \ref{compositionalitytheorem} we have 

\vspace{0.5mm}\noindent{\footnotesize
\begin{minipage}[b]{1\linewidth}  
\centering{\epsfig{figure=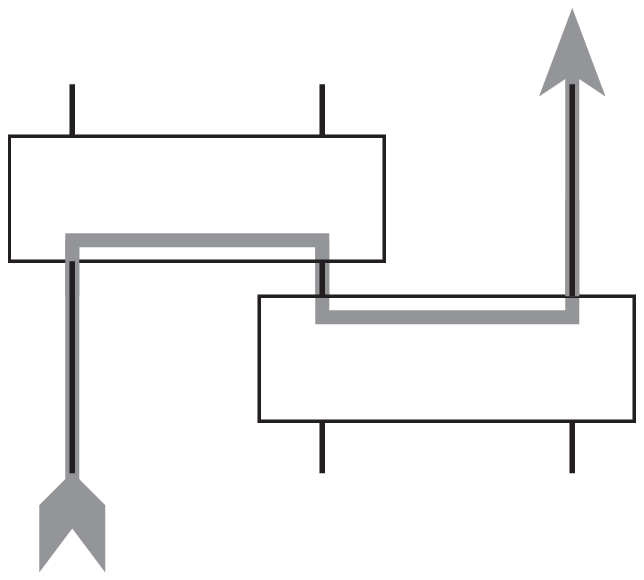,width=60pt}}   

\begin{picture}(160,0)   
\put(89,27){${\rm id}$} 
\put(66.5,44){$\!{\rm id}$}  
\put(49,19){$\phi$} 
\put(107,60){\small$\phi$}  
\put(186,42){\normalsize(3)}  
\end{picture}   
\end{minipage}}

\vspace{-3mm}\noindent
due to $({\rm id}\circ{\rm
id})(\phi)=\phi$.  When conceiving the first projector as
the preparation of an {EPR-pair} while tilting the tracks 
we indeed obtain `a' teleportation protocol.

\vspace{2.5mm}\noindent{\small
\begin{minipage}[b]{1\linewidth}  
\centering{\epsfig{figure=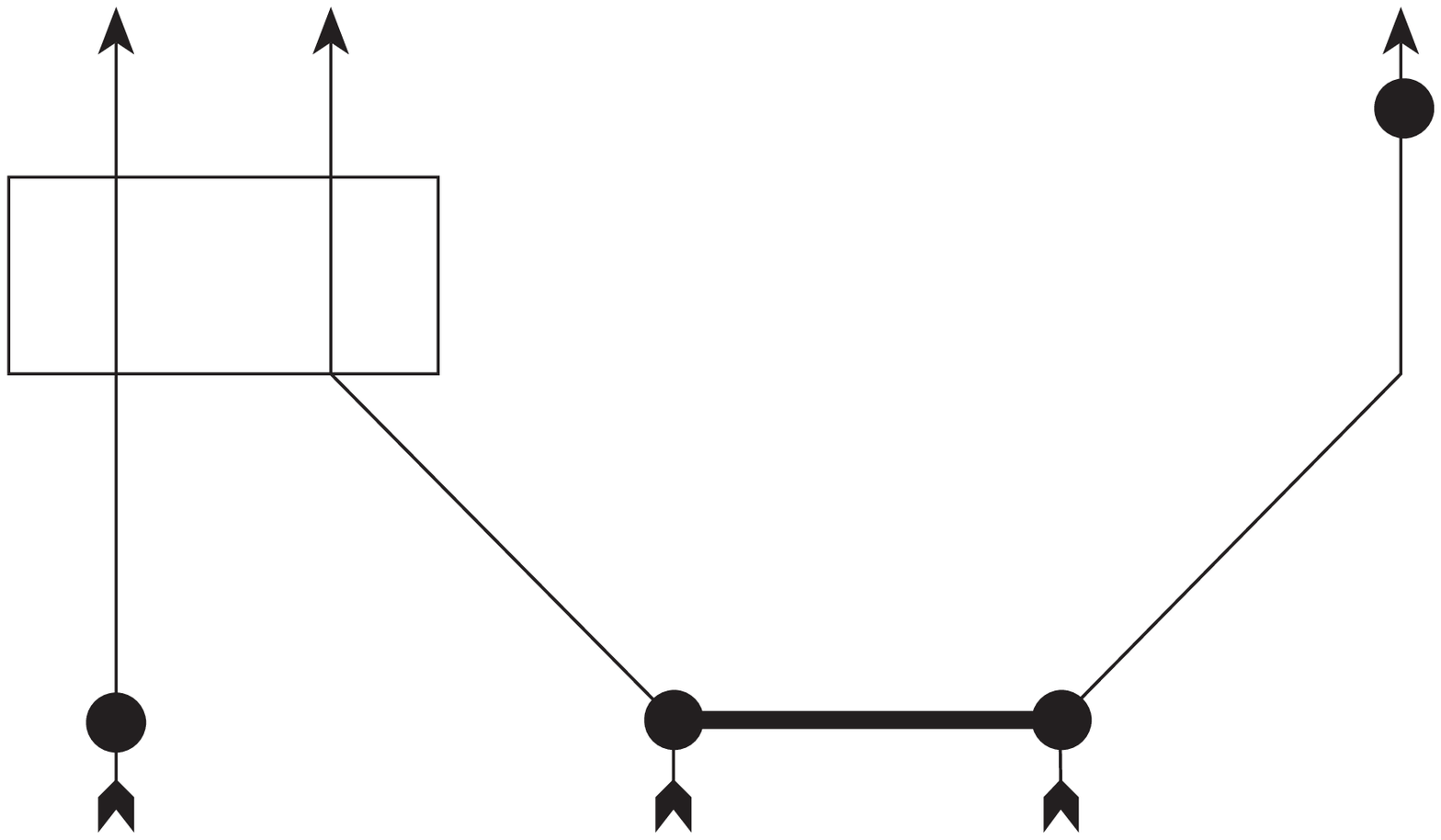,width=100pt}}     
  
\begin{picture}(120,0)   
\put(66.5,22){${\rm id}$} 
\put(21.5,48){${\rm id}$} 
\put(111.5,61){$\phi$}  
\put(9,18){$\phi$} 
\end{picture}  
\end{minipage}}

\vspace{-3mm}\noindent
However, the other projector has to `belong to a
measurement' e.g.~$M_{Bell}:=\{{\rm P}_{\rm id},{\rm P}_\pi,{\rm P}_{{\rm id}^*},{\rm
P}_{\pi^*}\}$.
Hence the above introduced protocol is
a {\it conditional\,} one. We want to make it {\it unconditional}.
\begin{definition}\em
Paths are {\it equivalent\,} iff for each input $\phi_{in}$ they
produce the same output $\phi_{out}$.
\end{definition}
\begin{corollary}\label{cor:Comperm1} 
For $U$ unitary and $g\circ U=U\circ g$ 

\vspace{1.5mm}\noindent{\footnotesize
\begin{minipage}[b]{1\linewidth}  
\centering{\epsfig{figure=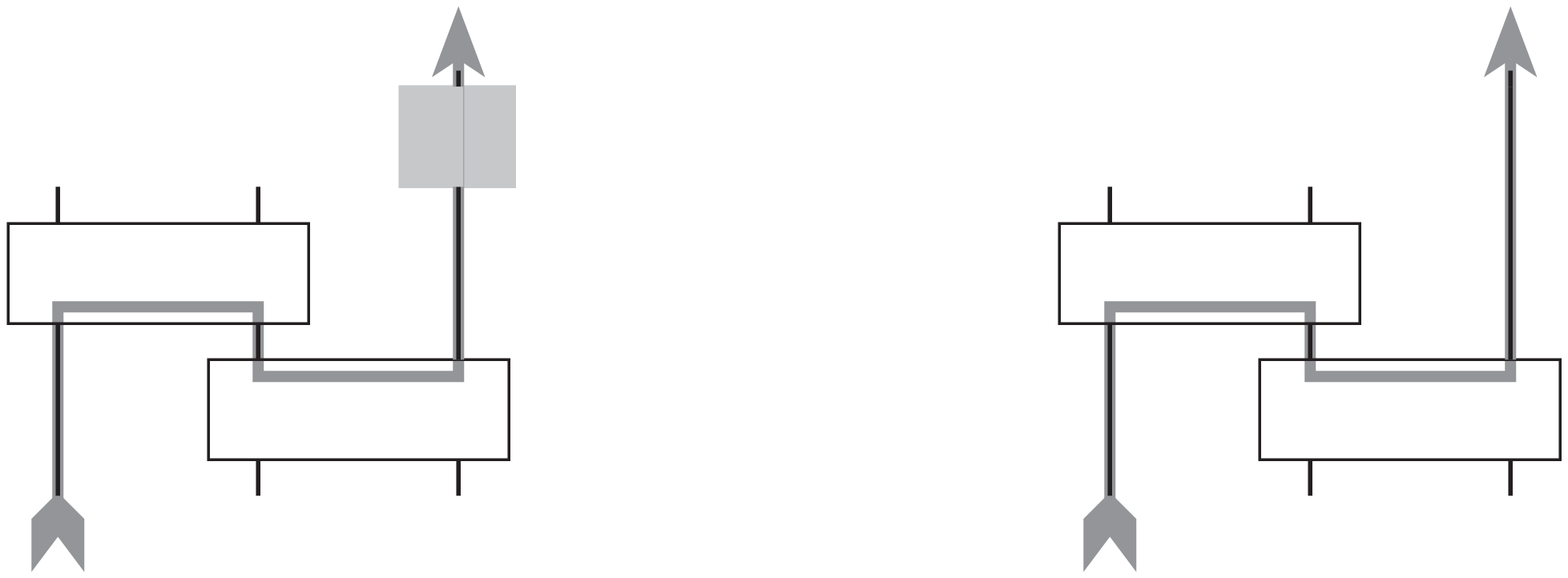,width=190.0pt}}     

\begin{picture}(290,0)   
\put(74,60){$U^\dagger$}  
\put(31.8,45.3){$U\circ f$}  
\put(63.5,27){$g$} 
\put(108,40){\normalsize and} 
\put(167.5,45.3){$f$}  
\put(192.5,27){$g$} 
\end{picture}   
\end{minipage}}

\vspace{-3mm}\noindent
are equivalent paths.
\end{corollary}
\bpf
Since 
$U^\dagger\!\circ g\circ (U\circ f)=g\circ f$
the result follows by Theorem \ref{compositionalitytheorem}. 
\hfill\endproof\newline 

\noindent  
Intuitively, one can move the box 
\raise-2.0pt\hbox{\vspace{2mm}\epsfig{figure=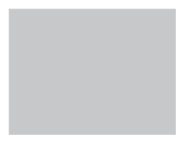,width=13.3pt}}\hspace{-3.8mm}{\small$U^\dagger$}\hspace{0.2mm}
along the path and permute it with projectors whose functional labels commute with $U$
(=\,commute with $U^\dagger$) until 
it gets annihilated by the $U$-factor of 
\raise-2.8pt\hbox{\vspace{2mm}\epsfig{figure=eProjector.eps,width=30pt}}\hspace{-8.9mm}{\small$U\circ
f$}\hspace{2.3mm}.
Applying Corollary \ref{cor:Comperm1} to 
\[
f,g:={\rm id}\quad \ \ {\rm and} \ \ \quad U\in\{{\rm id},\pi,{\rm id}^*,\pi^*\}\,,
\] 
since $\pi^\dagger=\pi$, $({\rm id}^*)^\dagger={\rm id}^*$ and 
$(\pi^*)^\dagger=-\pi^*\!$,
we obtain four conditional teleportation protocols 

\vspace{2mm}\noindent
\begin{minipage}[b]{1\linewidth}  
\centering{\epsfig{figure=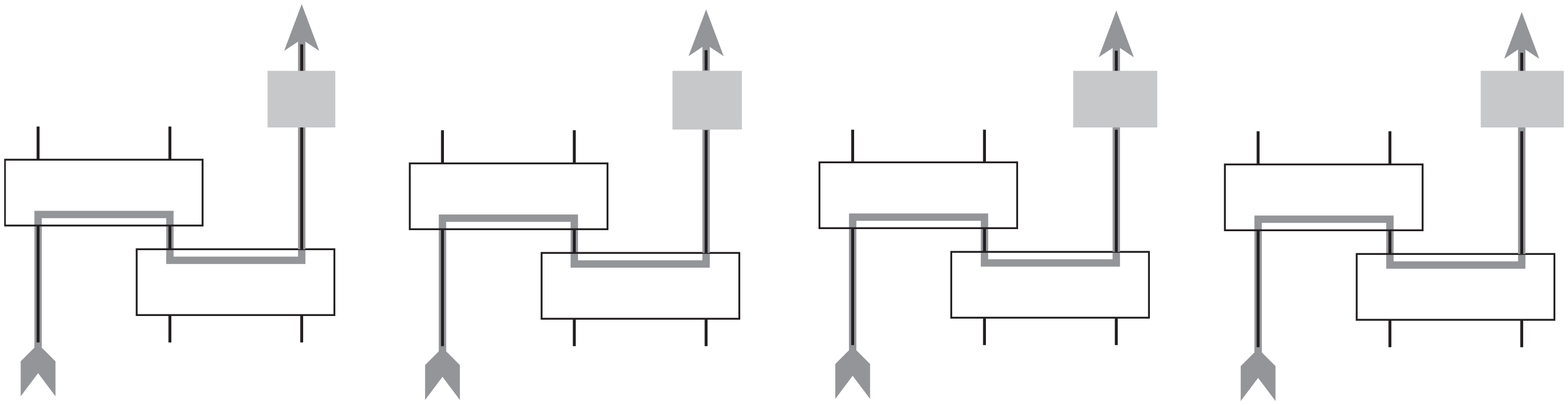,width=225pt}}   

\begin{picture}(210,0)   
\put(23,26){${\scriptstyle{\rm id}}$}   
\put(80,25.5){${\scriptstyle{\rm id}}$}  
\put(138,25.5){${\scriptstyle{\rm id}}$}  
\put(196,25.5){${\scriptstyle{\rm id}}$}  
\put(4.2,40.2){${\scriptstyle{\rm id}}$}  
\put(62.5,40){${\scriptstyle\pi}$}  
\put(119.7,40.1){${\scriptstyle{\rm id}^*}$}   
\put(177.5,39.8){${\scriptstyle\pi^*}$}  
\put(32.0,52.9){${\scriptstyle{\rm id}}$}  
\put(90.7,52.9){${\scriptstyle\pi}$}  
\put(147.2,52.9){${\scriptstyle{\rm id}^{\!*}}$} 
\put(204,52.9){{\tiny -}${\scriptstyle\pi^{\!*}}$}   
\end{picture}  
\end{minipage}  

\vspace{-3mm}\noindent
of which the one with $U:={\rm id}$ coincides with (3). 
These four together constitute an unconditional teleportation
protocol since they correspond to the four paths `from 
root to leaf' of the tree discussed in Section \ref{sec:class}, from
which then also the 2-bit classical channel emerges. 
%
%

\smallskip
In order to obtain the teleportation protocol as it
is found in the literature, observe that $\pi^*=\pi\circ id^*$, hence
\begin{center}{\small
\begin{tabular}{c|c|c|}
$\circ$ & ${\rm id}$ & ${\rm id}^*$ \\
\hline
${\rm id}$ & ${\rm id}$ & ${\rm id}^*$ \\
\hline
$\pi$ & $\pi$ & $\pi^*$ \\
\hline
\end{tabular}}
\end{center}
and thus we can factor --- with respect to composition of functional labels --- the 2-bit Bell-base
measurement in two 1-bit `virtual' measurements ($\vee$ stands for `or'):

\vspace{2mm}\hspace{-6.2mm}\noindent{\scriptsize
\begin{minipage}[b]{1\linewidth}  
\centering{\epsfig{figure=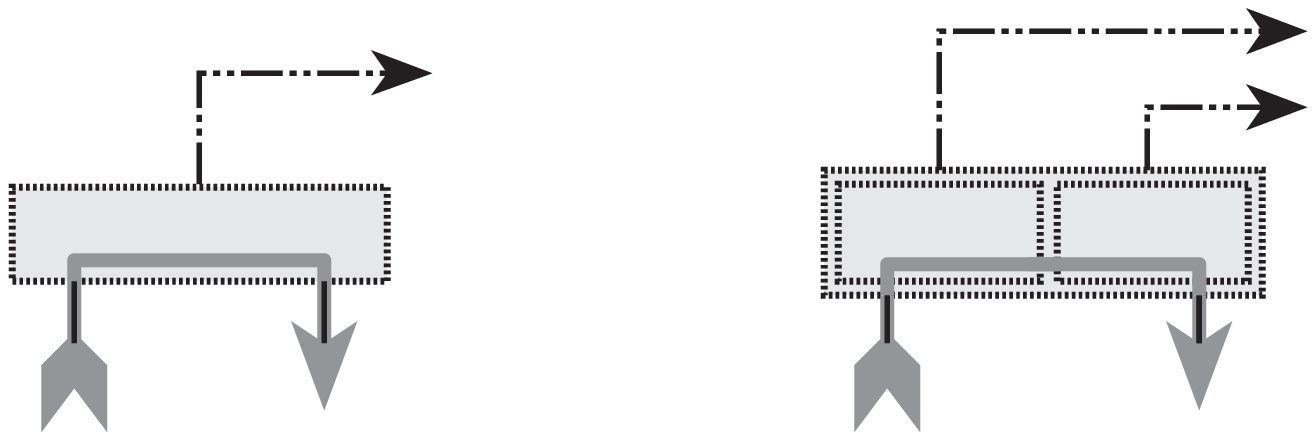,width=170pt}}       
    
\begin{picture}(0,0)      
\put(25.5,33.3){${{\rm id}\hspace{-1.8pt}\vee\hspace{-1.8pt}{\rm 
id}^{\hspace{-0.5pt}*}}$}   
\put(55,33.3){${{\rm id}\hspace{-1.5pt}\vee\hspace{-1.5pt}\pi}$}   
\put(-83.1,33.8){${{\rm id}\hspace{-1.5pt}\vee\hspace{-2pt}\pi\hspace{-1.5pt}
\vee\hspace{-1.5pt}{\rm id}^{\hspace{-0.5pt}*}\hspace{-2.5pt}\vee\hspace{-1.5pt}\pi^{\hspace{-0.5pt}*}}$}   
\put(-11,32){\Large$\simeq$} 
\put(-29,52.5){\footnotesize 2 bits}
\put(83.5,58){\footnotesize 1 bit} 
\put(83.5,48){\footnotesize 1 bit} 
\end{picture}  
\end{minipage}}

\vspace{-3.5mm}\noindent
Note that such a decomposition of $M_{Bell}$ does not exist with 
respect to $\otimes$ nor does it exist with respect to composition of 
projector actions.  All this results in

\vspace{1.5mm}\noindent
\begin{minipage}[b]{1\linewidth}  
\centering{\epsfig{figure=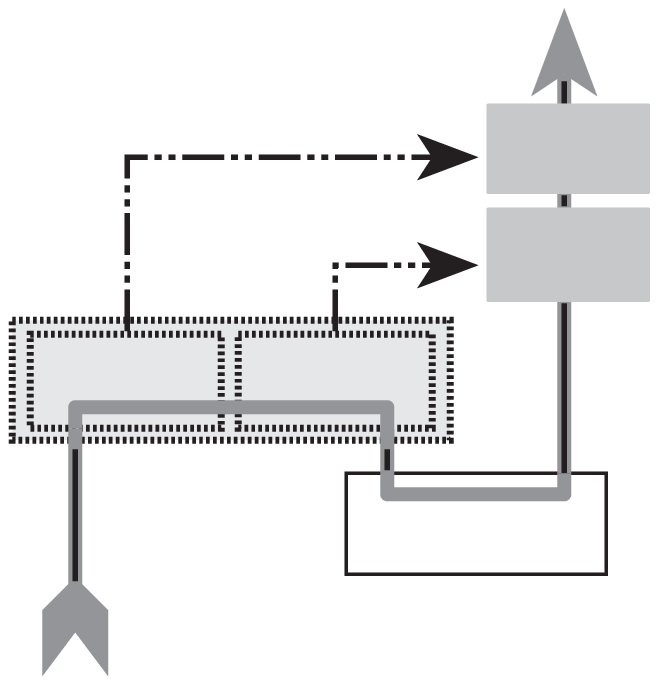,width=120pt}}  
    
\begin{picture}(230,0)     
\put(120,30.5){\small id}   
\put(62,55){${\scriptstyle{\rm id}\hspace{0.1pt}\vee\hspace{0.3pt}{\rm
id}^{\hspace{-0.5pt}*}}$}   
\put(95,55){${\scriptstyle{\rm id}\hspace{0.1pt}\vee\hspace{-0.1pt}\pi}$}   
\put(129.5,72){${\scriptstyle{\rm
id}\hspace{0pt}\vee\hspace{-0.1pt}\pi}$}   
\put(127.8,87){${\scriptstyle{\rm
id}\hspace{-0.3pt}\vee\hspace{0pt}{\rm
id}^{\hspace{-0.5pt}*}}$}   
\end{picture}  
\end{minipage} 

\vspace{-4mm}\noindent  
which is the standard teleportation protocol \cite{BBC}.

\smallskip
The aim of {\it logic
gate teleportation\,} \cite{Gottesman} is to teleport a state
and at the same time subject it to the action
of a gate $f$. By Theorem
\ref{compositionalitytheorem} we
evidently have

\vspace{2.5mm}\noindent{\footnotesize
\begin{minipage}[b]{1\linewidth}  
\centering{\epsfig{figure=eTeleport5.eps,width=60pt}}  

\begin{picture}(160,0)   
\put(89,26.8){$f$} 
\put(66.5,44){$\!{\rm id}$}  
\put(49,19){$\phi$} 
\put(107.8,60){\small$f(\phi)$}   
\end{picture}   
\end{minipage}}

\vspace{-3mm}\noindent
We make this protocol
unconditional analogously as we did it for
ordinary teleportation.   
\begin{corollary}\label{cor:Comperm}
For $U$ and $V$ unitary and $g\circ V=U\circ g$

\vspace{1.5mm}\noindent{\footnotesize
\begin{minipage}[b]{1\linewidth}   
\centering{\epsfig{figure=eTeleport2.eps,width=190pt}}    

\begin{picture}(290,0)   
\put(74,60){$U^\dagger$}  
\put(31.8,45.3){$V\circ f$}  
\put(63.5,27){$g$} 
\put(108,40){\normalsize and} 
\put(167.5,45.3){$f$}  
\put(192.5,27){$g$} 
\end{picture}   
\end{minipage}}

\vspace{-3mm}\noindent
are equivalent paths.
\end{corollary}
\bpf
Analogous to that of Corollary \ref{cor:Comperm}.  
\hfill\endproof\newline

\noindent  
We apply the above to the case  
\[
f:={\rm id}\otimes{\rm id}\quad\quad\ \ {\rm and}\quad\quad\ \ 
g:=\textsc{\footnotesize CNOT} 
\]
that is, the first projector is now to be conceived as
the preparation of the state
\[
\Psi_{\textsc{\tiny CNOT}}=
|00\rangle
\otimes
|00\rangle
+ 
|01\rangle
\otimes
|01\rangle
+
|10\rangle
\otimes
|11\rangle
+  
|11\rangle
\otimes
|10\rangle\,.
\]
Let $\Psi_f$ be defined either by $f\stackrel{\rm L}{\simeq}\Psi_f$ or $f\stackrel{\rm aL}{\simeq}\Psi_f$.
\begin{proposition}\label{lm:tenslabel}
$\Psi_{f\otimes g}=\Psi_{f}\otimes\Psi_{g}$\,{\rm ;}
${\rm P}_{f\otimes g}={\rm P}_{f}\otimes{\rm P}_{g}$.
\end{proposition} 
\bpf 
The first claim is verified straightforwardly. Hence 
${\rm P}_{f\otimes g}\equiv{\rm P}_{\Psi_{f\otimes g}}={\rm 
P}_{\Psi_{f}\otimes\Psi_{g}}={\rm P}_{\Psi_f}\otimes{\rm
P}_{\Psi_g}\equiv{\rm P}_{f}\otimes{\rm P}_{g}$ 
what completes the proof.
\hfill\endproof\newline

\noindent  
Hence we can factor the 4-qubit measurement to which the second projector belongs in two Bell-base
measurements, that is, we set
\[
V\in\left\{U_1\otimes U_2\bigm| U_1,U_2\in \{{\rm id},\pi,{\rm id}^*,\pi^*\}\right\}\,.
\] 
The resulting protocol 

\vspace{-1.5pt}\noindent
\begin{minipage}[b]{1\linewidth}  
\centering{\epsfig{figure=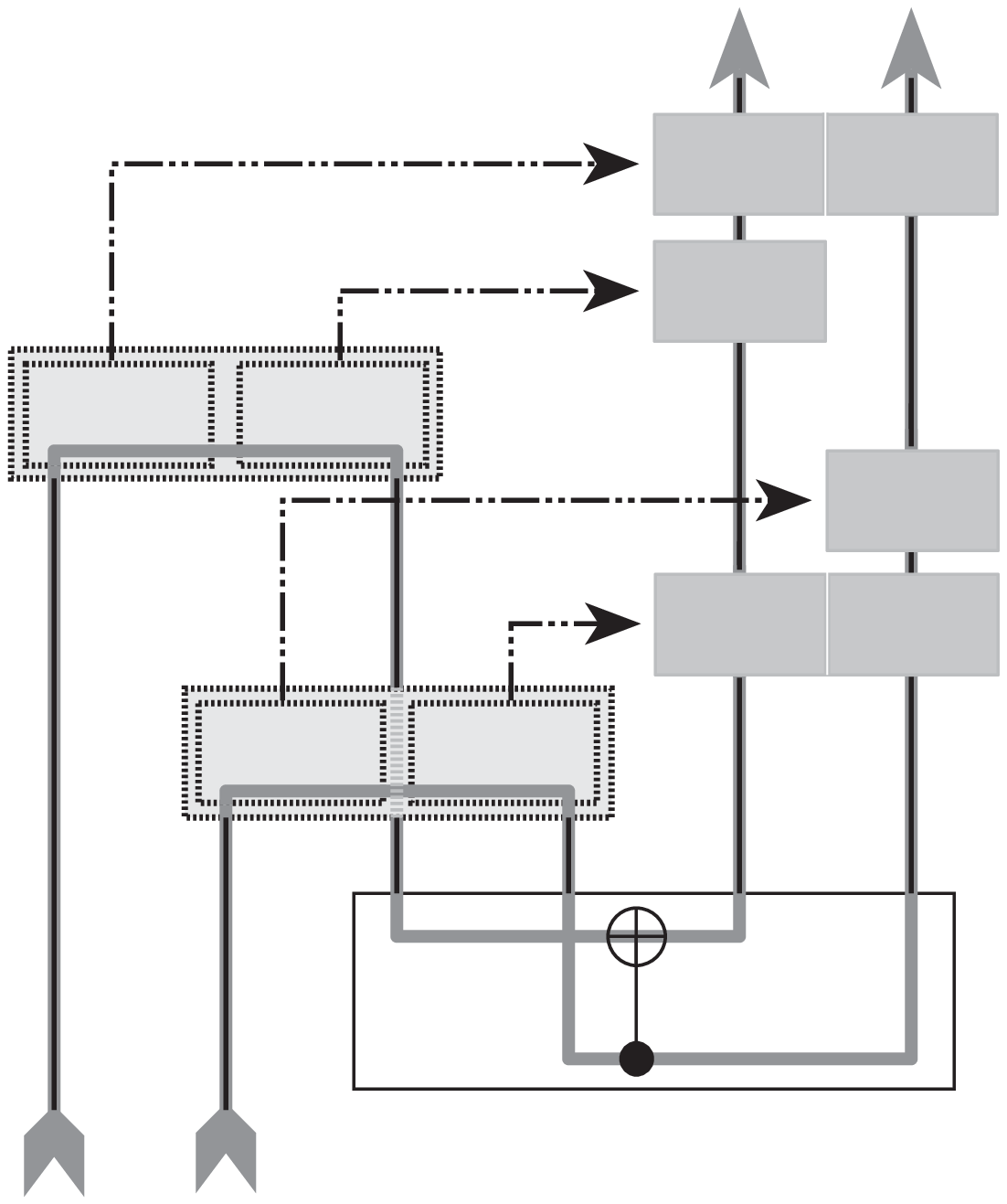,width=170pt}}  
    
\begin{picture}(170,0)     
\put(103,43){\textsc{\footnotesize CNOT}}    
\put(35,82){${\scriptstyle{\rm
id}\hspace{0.1pt}\vee\hspace{0pt}{\rm id}^{\hspace{-0.5pt}*}}$}   
\put(71,82){${\scriptstyle{\rm
id}\hspace{0.1pt}\vee\hspace{-0.1pt}\pi}$}   
\put(8,135){${\scriptstyle{\rm
id}\hspace{0.1pt}\vee\hspace{0pt}{\rm id}^{\hspace{-0.5pt}*}}$}   
\put(44,135){${\scriptstyle{\rm
id}\hspace{0.1pt}\vee\hspace{-0.1pt}\pi}$}   
\put(133.5,120.5){${\scriptstyle{\rm
id}\hspace{0.1pt}\vee\hspace{0pt}{\rm id}^{\hspace{-0.5pt}*}}$}   
\put(133.5,174){${\scriptstyle{\rm
id}\hspace{0.1pt}\vee\hspace{0pt}{\rm id}^{\hspace{-0.5pt}*}}$}   
\put(105.5,174){${\scriptstyle{\rm
id}\hspace{0.1pt}\vee\hspace{0pt}{\rm id}^{\hspace{-0.5pt}*}}$}   
\put(135.5,101.5){${\scriptstyle{\rm
id}\hspace{0.1pt}\vee\hspace{-0.1pt}\pi}$}   
\put(108,101.5){${\scriptstyle{\rm
id}\hspace{0.1pt}\vee\hspace{-0.1pt}\pi}$}   
\put(108,154){${\scriptstyle{\rm
id}\hspace{0.1pt}\vee\hspace{-0.1pt}\pi}$}    
\end{picture}  
\end{minipage}  

\vspace{-4mm}\noindent  
is the one to be found in \cite{Gottesman} --- recall that
$U^\dagger$ factors as a tensor since
\textsc{\footnotesize CNOT} is a member of the Clifford
group.

\smallskip
Our last example in this section involves the passage {\it from sequential to parallel composition of logic
gates}.   Due to the accumulation of inaccuracies in sequential composition \cite{Preskill}
it would be desirable to have a fault-tolerant parallel alternative.  This would for example be useful if we have a
limited set of available gates from which we want to generate more general ones e.g.~generating all Clifford
group gates from
\textsc{\footnotesize CNOT} gates, Hadamard gates and phase gates via tensor and composition. By Theorem 
\ref{compositionalitytheorem} the network

\vspace{-0.5mm}\noindent{\footnotesize
\begin{minipage}[b]{1\linewidth}  
\centering{\epsfig{figure=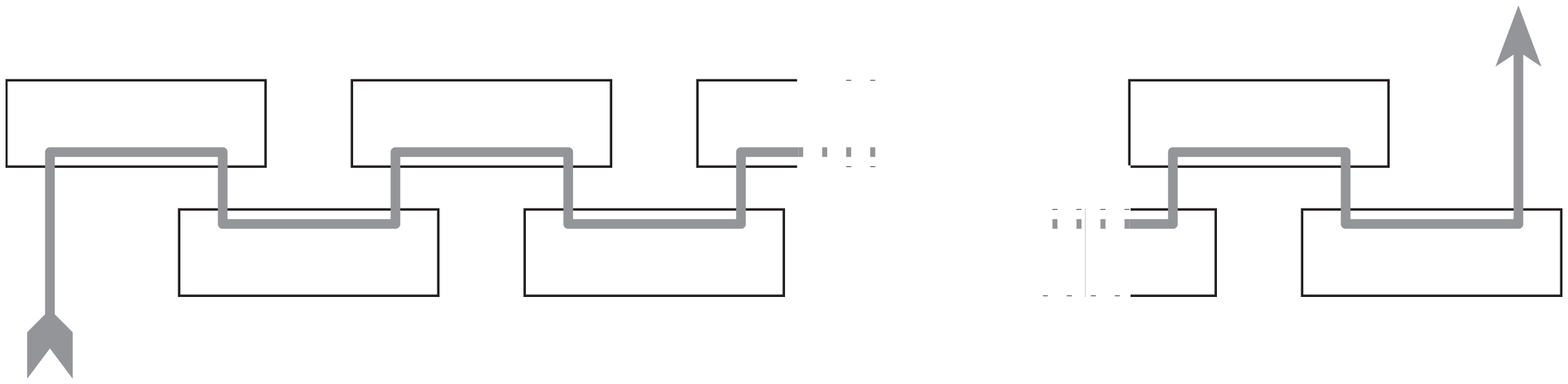,width=210pt}}      

\begin{picture}(210,0)      
\put(15,43){id} 
\put(38,23.5){$f_1$}  
\put(62,43){id}  
\put(85,23.5){$f_2$} 
\put(166,43){id}  
\put(189,23.5){$f_m$} 
\end{picture}  
\end{minipage}}
 
\vspace{-3mm}\noindent
realizes the composite $f_m\circ\ldots\circ f_1$ conditionally.
Again this protocol can be made unconditional --- an
algorithm which captures the general case can be found in 
\cite{RR} \S3.4.  Note that by Theorem 
\ref{compositionalitytheorem} it suffices to make unitary
corrections only at the end of the path \cite{RR} \S3.4.  

\section{Entanglement swapping}\label{sec:outputonly}

By Theorem \ref{compositionalitytheorem} we have 

\vspace{-2.5mm}\noindent{\footnotesize
\begin{minipage}[b]{1\linewidth}   

\vspace{4mm}\noindent 
\centering{\epsfig{figure=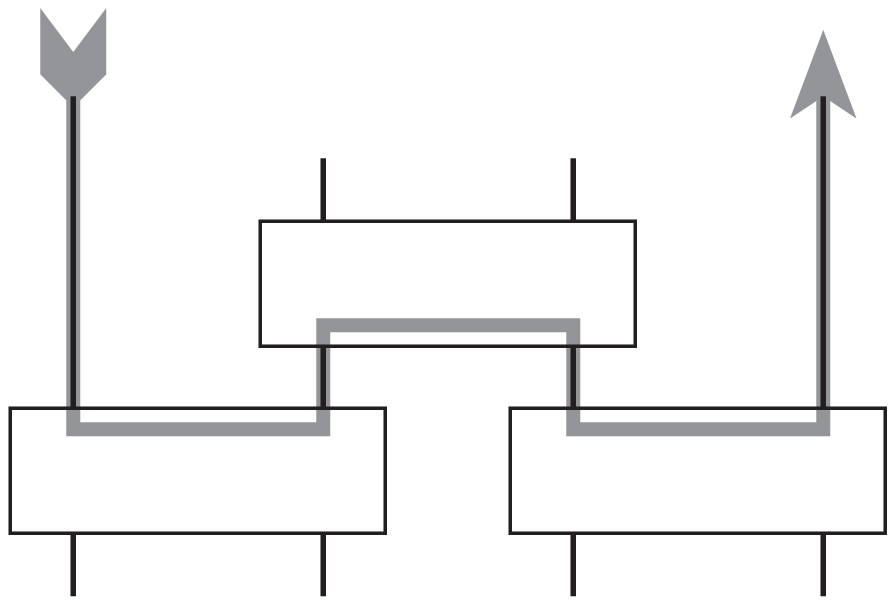,width=100pt}}  
 
\begin{picture}(100,0)      
\put(-2,56){$\phi_{in}$}
\put(94,56){$(h\circ g\circ f)(\phi_{in})$}   
\put(47.8,41.7){$g$}
\put(22.2,19.5){$f$} 
\put(73.2,19.5){$h$}
\put(8.5,2.5){${\cal H}_1$}
\put(33.5,2.5){${\cal H}_2$} 
\put(58.5,2.5){${\cal H}_3$}
\put(83.5,2.5){${\cal H}_4$}
\end{picture}
\end{minipage}} 
 
\vspace{-0.5mm}\noindent
However, Theorem \ref{compositionalitytheorem}
assumes $\phi_{in}$ to be a pure factor while it is part of the 
output $\Psi_{\bf out}$ of the network. This fact
constraints the network by requiring that
\[
h\circ g\circ f\,\,\stackrel{\rm aL}{\simeq}\,\,\phi_{in}\otimes\phi_{out} 
\]
for some $\phi_{in}$ and $\phi_{out}$ i.e.~the state labeled by $h\circ g\circ f$ has to be
{\it disentangled} --- which is equivalent to the range of $h\circ g\circ f$ 
being one-dimensional \cite{RR}~\S 5.3. 

\smallskip
Using Lemma \ref{lm:compos} this pathology can be 
overcome by conceiving the output state of the bipartite subsystem described in
${\cal H}_1\otimes{\cal H}_4$ not as a {\it pair\,} $(\phi_{in},\phi_{out})$ but as a {\it
function\,}
$\varphi:{\cal H}_1\!\lto\!{\cal H}_4$ which relates any input $\phi_{in}\in{\cal H}_1$ to an output
$\phi_{out}:=\varphi(\phi_{in})\in{\cal H}_4$.  Hence we conceive the above network as producing a function
\[ 
\varphi:=h\circ g\circ f\,\,\stackrel{\rm aL}{\simeq}\,\,\Psi_\varphi
\]  
where $\Psi_{\bf out}=\Psi_\varphi\otimes\Psi_g$ with 
\[
\Psi_\varphi\in{\cal H}_1\otimes{\cal H}_4
\quad{\rm and}\quad
g\stackrel{\rm aL}{\simeq}\Psi_g\in{\cal H}_2\otimes{\cal H}_3\,.
\] 
To such a function produced by a network we can provide
an input via a unipartite projector. The generic example (which
can be easily verified) is 

\vspace{1.5mm}\noindent{\footnotesize  
\begin{minipage}[b]{1\linewidth}  
\centering{\epsfig{figure=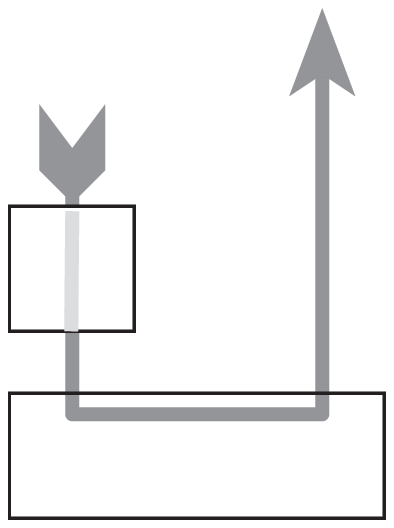,width=40pt}}    

\begin{picture}(40,0)       
\put(17,13){$f$} 
\put(1.3,34.5){${\phi}_{in}$}
\put(37.5,51){$\phi_{out}=f(\phi_{in})$}  
\end{picture}
\end{minipage}}   

\vspace{-3mm}\noindent
One can then conceive 
\raise-2.6pt\hbox{\vspace{2mm}\epsfig{figure=eProjector.eps,width=28pt}}\hspace{-5.5mm}{\footnotesize$f$}\hspace{4.9mm}
as a $\lambda$-{\it term\,} $\lambda\phi.f\phi$ \cite{Barendrecht} and the process of
providing it with an input via a unipartite projector embodies the $\beta$-{\it reduction\,} \cite{Barendrecht}
\[
(\lambda\phi.f\phi)\phi_{in}\stackrel{\beta}{=}f(\phi_{in})\,. 
\]
As we will see below we can `feed' such a function at its turn as an input of {\it function
type\,} in another network. This view carries over to the interpretation of multipartite 
entanglement where it becomes crucial.

\smallskip
The entanglement swapping protocol \cite{Swap} can now be derived analogously as the teleportation
protocol by setting
$f=g=h:={\rm id}$ in the above.  For this particular case Lemma \ref{lm:compos} becomes

\vspace{3.5mm}\noindent{\footnotesize 
\begin{minipage}[b]{1\linewidth}  
\centering{\epsfig{figure=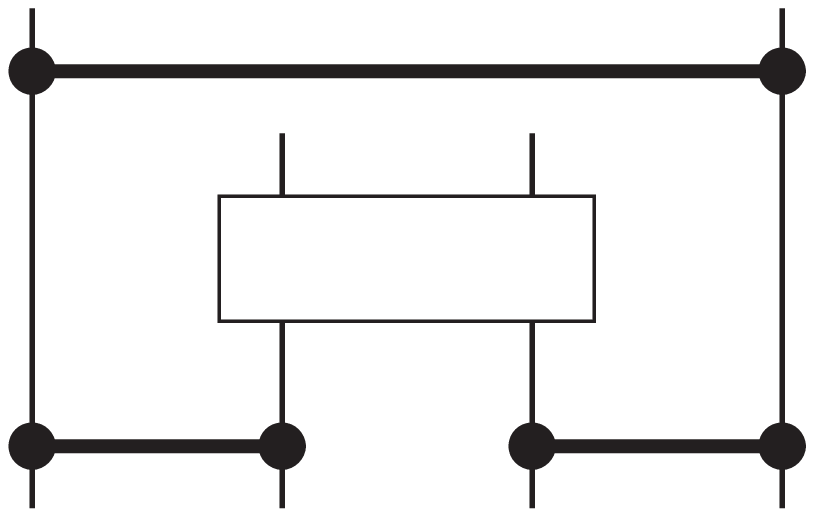,width=100pt}}  

\begin{picture}(100,0)       
\put(23,58){${\rm id}\circ {\rm id}\circ {\rm id}={\rm id}$}  
\put(22.5,20){id}
\put(72,20){id}  
\put(47.8,33){id} 
\end{picture}
\end{minipage}}

\vspace{-1.5mm}\noindent
Details can be found in  \cite{RR} \S 6.2. 

\section{Multipartite entanglement}

The passage from states to functions as inputs and outputs enables to extend our functional interpretation of
bipartite entanglement to one for multipartite entanglement.  In general this involves {\it higher order functions}
and hence the use of denotational tools from modern logic and proof theory such as $\lambda$-calculus
\cite{Abr,Barendrecht}.

\smallskip
Whereas (due to commutativity of $-\otimes-$) a bipartite tensor ${\cal H}_1\otimes{\cal H}_2$ admits   
interpretation as a function either of type ${\cal
H}_1\!\lto\!{\cal H}_2$ or of type ${\cal H}_1\!\rto\!{\cal H}_2$,  a 
tripartite tensor (due to associativity of $-\otimes-$)
admits interpretation  as a function of a type within the union of two (qualitatively different)
families of types namely
\[
{\cal H}_i\lto({\cal H}_j\lto{\cal H}_k)\ \ \ \  {\rm and}\ \ \ \ ({\cal H}_i\lto{\cal H}_j)\lto{\cal H}_k\,.
\]
Explicitly, given 
\[
\sum_{\alpha\beta}M_{\alpha\beta\gamma}\cdot {e}_{\alpha}^{(1)}\otimes{e}_{\beta}^{(2)}\otimes{e}_{\gamma}^{(3)}
\,\in\,
{\cal H}_1\otimes{\cal H}_2\otimes{\cal H}_3
\]
we respectively obtain  
\beqa
f_1\!\!\!\!\!&:&\!\!\!\!\!{\cal H}_1\lto({\cal H}_2\lto{\cal H}_3)\\
\!\!\!\!\!&::&\!\!\!\!\!\sum_\alpha\psi_\alpha\cdot {e}_{\alpha}^{(1)}
\mapsto
\sum_{\beta\gamma}\Bigl(\sum_{\alpha}\bar{\psi}_{\alpha}M_{\alpha\beta\gamma}\Bigr)
\langle-\mid{e}_{\beta}^{(2)}\rangle\cdot {e}_{\gamma}^{(3)}
\eeqa
and
\beqa
f_2\!\!\!\!\!&:&\!\!\!\!\!({\cal H}_1\lto{\cal H}_2)\lto{\cal H}_3\\
\!\!\!\!\!&::&\!\!\!\!\!\sum_{\alpha\beta}m_{\alpha\beta}\langle-\mid{e}_{\alpha}^{(1)}\rangle\cdot
{e}_{\beta}^{(2)}
\mapsto
\sum_\gamma\Bigl(\sum_{\alpha\beta}\bar{m}_{\alpha\beta}M_{\alpha\beta\gamma}\Bigr)\cdot e_{\gamma}^{(3)}
\eeqa
as the corresponding functions --- the complex conjugation of the coefficients
$\bar{\psi}_{\alpha}$ and $\bar{m}_{\alpha\beta}$ is due to the anti-linearity of the maps. The
appropriate choice of an interpretation for a tripartite projector depends on the {\it context}
i.e.~the configuration of the whole network to which it belongs. A {\it first order function\,}
$f_1$ enables interpretation in a configuration such as 
  
\vspace{-2.5mm}\noindent{\scriptsize  
\begin{minipage}[b]{1\linewidth}  

\bigskip\noindent 
\centering{\epsfig{figure=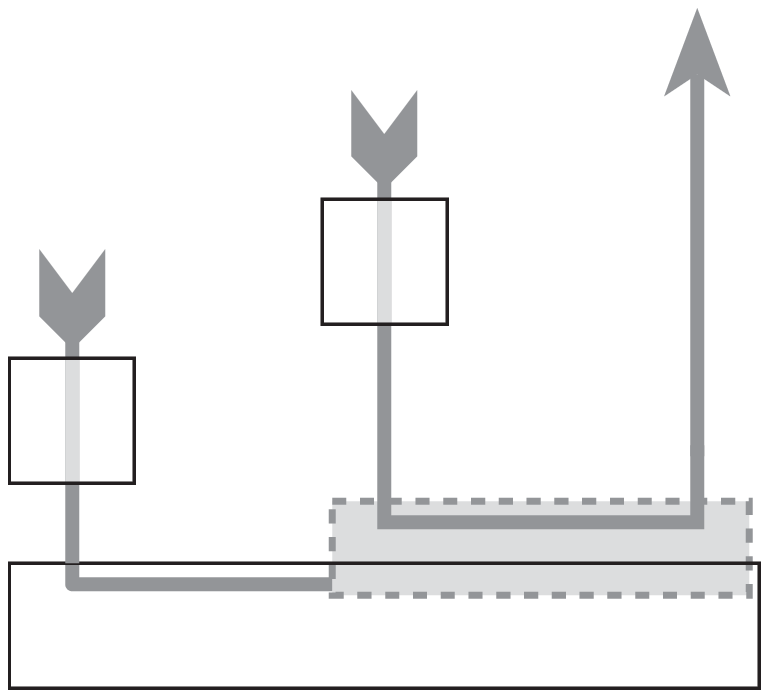,width=80pt}}     
 
\begin{picture}(80,0)      
\put(5.0,12){$f_1:{\cal H}_1\!\lto\!({\cal
H}_2\!\lto\!{\cal H}_3)$} 
\put(3.5,35){$\phi_1$} 
\put(36.5,52){$\phi_2$} 
\put(76,66.5){\small$\,\phi_{out}\!\!=(f(\phi_1))(\phi_2)$}  
\end{picture}
\end{minipage}}    

\vspace{-2mm}\noindent
One can think of this tripartite projector as producing
a bipartite one at its `output'. A 
{\it second order function\,}
$f_2$ --- recall that a definite integral is an example of a second order function --- enables
interpretation in the configuration 
 
\vspace{-1.5mm}\noindent{\scriptsize
\begin{minipage}[b]{1\linewidth}  

\noindent  
\centering{\epsfig{figure=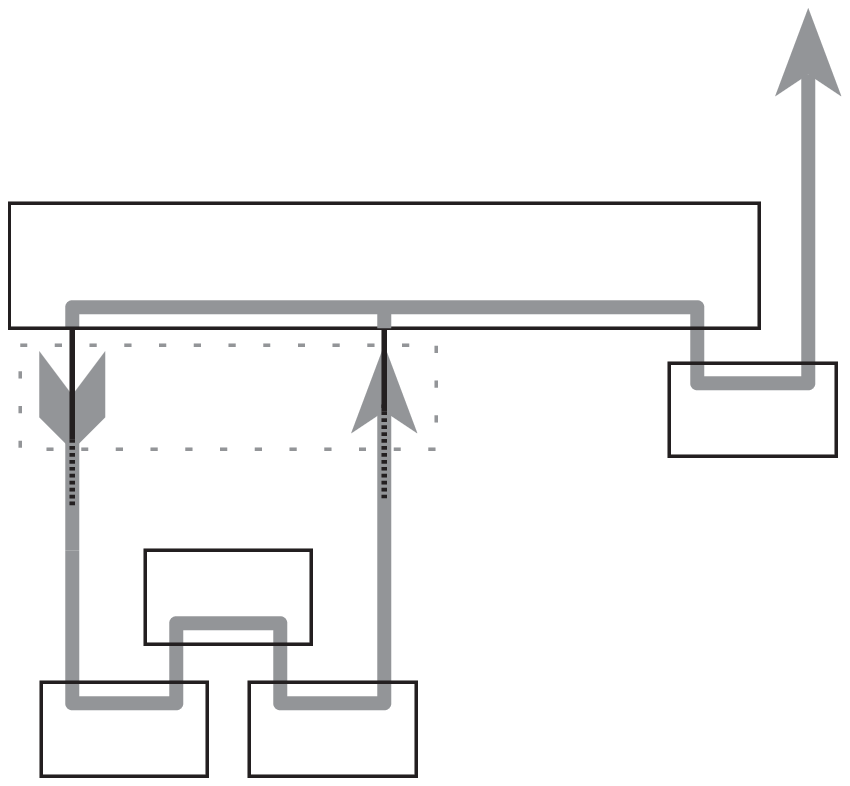,width=90pt}}  
 
\begin{picture}(90,0)      
\put(5,63.2){$f_2:({\cal H}_1\!\lto\!{\cal H}_2)\!\lto\!{\cal
H}_3$}   
\put(23,47.5){\small$g$} 
\put(88.7,76.5){\small$\,\phi_{out}\!\!=f(g)$}  
\end{picture}
\end{minipage}}   

\vspace{-2mm}
We illustrate this in an example --- we will not provide an analogue to Theorem
\ref{compositionalitytheorem} for the  multipartite case since even its formulation requires advanced
denotational tools. Consider the following configuration.

\vspace{2mm}\noindent
\hspace{2.5mm}\begin{minipage}[b]{1\linewidth}  
\centering{\epsfig{figure=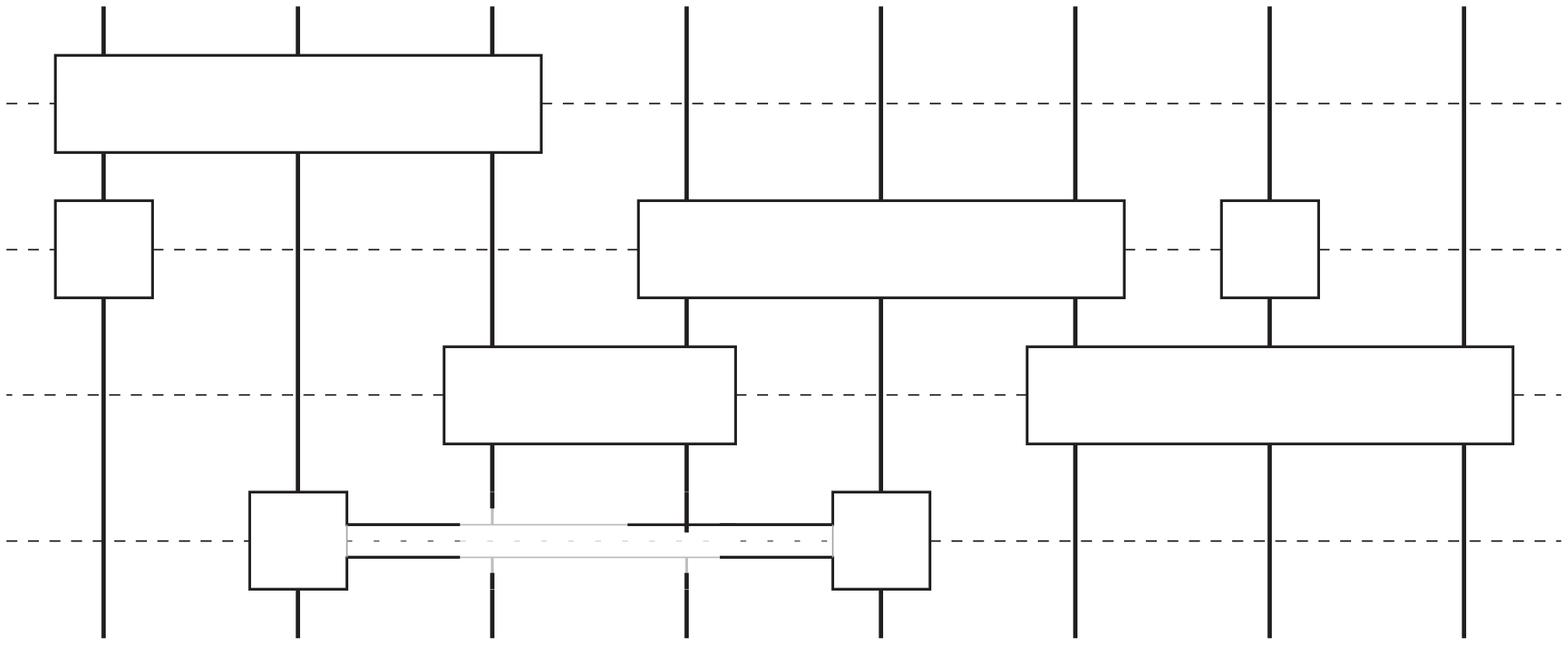,width=213.3pt}} 

{\scriptsize 
\begin{picture}(213.3,0)    
\put(-8,25){$1$}
\put(-8,45){$2$}
\put(-8,65){$3$}
\put(-8,85){$4$} 
\put(10,4){${\cal H}_1$}
\put(36.6,4){${\cal H}_2$}
\put(63.3,4){${\cal H}_3$}
\put(90,4){${\cal H}_4$}
\put(116.6,4){${\cal H}_5$}
\put(142.4,4){${\cal H}_6$}
\put(169.4,4){${\cal H}_7$}
\put(196,4){${\cal H}_8$}
\put(10.0,84){$(M^1_{\alpha_1\alpha_2\alpha_3})_{\alpha_1\alpha_2\alpha_3}$} 
\put(89.7,64){$(M^2_{\alpha_4\alpha_5\alpha_6})_{\alpha_4\alpha_5\alpha_6}$} 
\put(142.7,44){$(M^3_{\alpha_6\alpha_7\alpha_8})_{\alpha_6\alpha_7\alpha_8}$}  
\put(63,44){\tiny$(m^1_{\alpha_{\!3}\!\alpha_{\!4}}\!)_{\alpha_{\!3}\!\alpha_{\!4}}$} 
\put(63.8,25){\tiny$(m^2_{\alpha_{\!2}\!\alpha_{\!5}}\!)_{\alpha_{\!2}\!\alpha_{\!5}}$}  
\put(8.5,64.4){$\ \phi_1$}   
\put(168.5,64.4){$\ \phi_2$} 
\end{picture}
}
\end{minipage}

\vspace{0mm}\noindent  
For `good' types we can draw a `compound' path.

{\tiny 
\vspace{2mm}\noindent
\begin{minipage}[b]{1\linewidth}  
\centering{\epsfig{figure=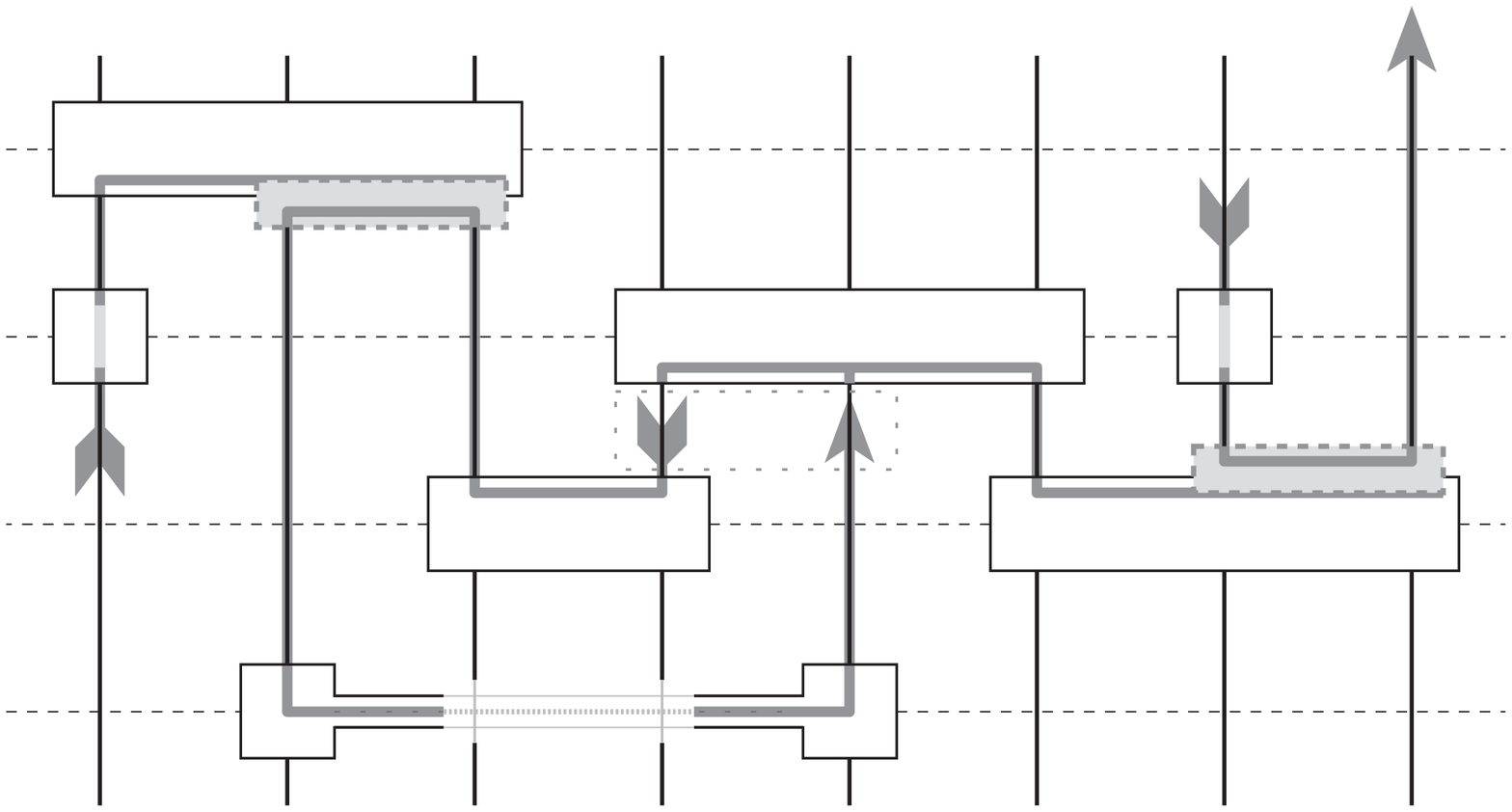,width=213.3pt}}  

\begin{picture}(213.3,0)   
\put(10,2){${\cal H}_1$}
\put(36.6,2){${\cal H}_2$}
\put(63.3,2){${\cal H}_3$} 
\put(90,2){${\cal H}_4$}
\put(116.6,2){${\cal H}_5$} 
\put(142.4,2){${\cal H}_6$}
\put(169.4,2){${\cal H}_7$} 
\put(196,2){${\cal H}_8$}
\put(11,100){$f_1\!\!:\!{\cal H}_1\!\lto\!(\!{\cal H}_2\!\lto\!{\cal H}_3\!)$}
\put(91,73){$f_2\!\!:\!(\!{\cal H}_4\!\lto\!{\cal H}_5\!)\!\lto\!{\cal H}_6$}
\put(144,44){$f_3\!\!:\!{\cal H}_6\!\lto\!(\!{\cal H}_7\!\lto\!{\cal
H}_8\!)$}
\put(63,44){$g_1^\dagger\!\!:\!{\cal H}_3\!\!\looparrowleft\!{\cal
H}_4$}
\put(62.9,19.5){$g_2\!:\!{\cal H}_2\!\lto\!{\cal H}_5$}
\put(11,72.5){$\phi_1$}  
\put(170,72.5){$\phi_2$} 
\put(201.5,104){\footnotesize $?[\phi_{out}]$}   
\end{picture}
\end{minipage}
}

\vspace{-3mm}\noindent 
If a multipartite analogue to Theorem \ref{compositionalitytheorem} truly holds one would obtain
\[
\phi_{out}=(f_3\circ f_2)(g_2\circ (f_1(\phi_1))^\dagger\circ
g_1^\dagger)(\phi_2)\,.
\]
Hence in terms of matrices we predict $\phi^{out}_{\alpha_8}$ to be
\[
\sum_{\alpha_1\ldots\alpha_7}
\hspace{-2mm}
\bar{\phi}^{2}_{\alpha_7}m^1_{\alpha_3\alpha_4}{\phi}^{1}_{\alpha_1}
\bar{M}^1_{\alpha_1\alpha_2\alpha_3}m^2_{\alpha_2\alpha_5}
\bar{M}^2_{\alpha_4\alpha_5\alpha_6}{M}^3_{\alpha_6\alpha_7\alpha_8}.
\]
To verify this we explicitly calculate $\phi^{out}_{\alpha_8}$.
Set
\[
\Psi^\tau=\sum_{i_1\ldots i_8}\Psi^\tau_{i_1\ldots i_8}\cdot
e_{i_1}^{(1)}\otimes\ldots\otimes e_{i_8}^{(8)}
\]
where $\Psi^0$ is the (essentially arbitrary) input of the 
network and $\Psi^\tau$ for
$\tau\in\{1,2,3,4\}$ is the state at time $\tau+\epsilon$. For
$I\subseteq\{1,\ldots,8\}$ and $I^{{}^c}\!:=\{1,\ldots,8\}\setminus
I$ let ${\rm P}_\Phi^{I}$ stipulate that this projector projects 
on the subspace
\[
\Phi\otimes\bigotimes_{i\in I^{{}^c}\!\!}{\cal H}_i  
\quad{\rm for\ some}\quad 
\Phi\in \bigotimes_{i\in I}{\cal H}_i\,.
\]
\begin{lemma}\label{projectoractionlemma}
If
$\Psi^{\tau}={\rm P}_\Phi^{I}(\Psi^{\tau-1})$ 
then 
\[
\Psi^{\tau}_{i_1\ldots i_8}=
\!\!\sum_{j_\alpha\mid\alpha\in I}\!\!
\Psi^{\tau-1}_{i_1\ldots i_8[j_\alpha/i_\alpha\mid\alpha\in
I]}\bar{\Phi}_{(j_\alpha\mid\alpha\in
I)}\Phi_{(i_\alpha\mid\alpha\in I)}
\]
where 
$i_1\ldots i_8[j_\alpha/i_\alpha\mid\alpha\in I]$ 
denotes that for $\alpha\in I$ we substitute the index 
$i_\alpha$ by the index $j_\alpha$ which ranges
over the same values as $i_\alpha$.
\end{lemma}
\bpf 
Straightforward verification or see \cite{RR} \S 6.4. 
\hfill\endproof\newline

\noindent 
Using Lemma \ref{projectoractionlemma} one verifies that the resulting state $\Psi^4_{i_1\ldots i_8}$ factors into
five components, one in which no index in 
$\{i_1,\ldots,i_8\}$ appears, three with indices in $\{i_1,\ldots,i_7\}$ and one which contains the index
$i_8$ namely
\[
\hspace{-4mm}
\sum_{
\begin{array}{c}
{\scriptstyle l_4l_5l_6l_7}\vspace{-1mm}\\ 
{\scriptstyle m_1m_2m_3}
\end{array}
}\hspace{-4mm}
m^2_{m_2l_5}m^1_{m_3l_4}M^3_{l_6l_7i_8}\phi^{1}_{m_1}
\bar{M}^2_{l_4l_5l_6}\bar{\phi}^{2}_{l_7}\bar{M}^1_{m_1m_2m_3}\,.  
\] 
Substituting the indices
$m_1$, $m_2$, $m_3$, $l_4$, $l_5$, $l_6$, $l_7$, $i_8$ by $\alpha_1,\ldots,\alpha_8$   
we exactly obtain our prediction for $\phi^{out}_{\alpha_8}$.

\smallskip
It should be clear from our discussion of multipartite entanglement 
that, provided we have an appropriate entangled state involving a
sufficient number of qubits, we can implement arbitrary
(linear) $\lambda$-terms \cite{Abr,Barendrecht}.

\section{Discussion}  

For a unitary operation $U:{\cal H}\to{\cal H}$ there is a {\it flow of information\,}
from the input to the output of $U$ in the sense that
for an input state $\phi$ the output $U(\phi)$ fully depends on $\phi$.  

\vspace{1.0mm}\noindent
\begin{minipage}[b]{1\linewidth}   
\centering{\epsfig{figure=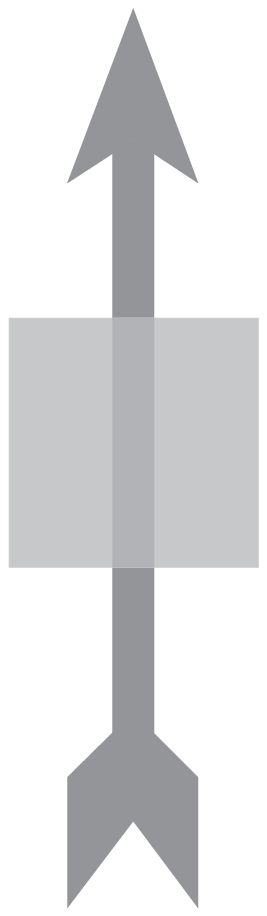,width=18pt}}   

\begin{picture}(18,0)   
\put(-42,16){\sf input:}  
\put(18,15){\large$\phi$}  
\put(5.5,41.5){\large$U$}  
\put(-42,67){\sf output:}  
\put(18,66){\large$U(\phi)$}  
\end{picture}   
\end{minipage}

\vspace{-4.0mm}\noindent 
How does a projector ${\rm P}_\psi$ act
on states?   After renormalization and provided that 
$\langle\phi\!\mid\! \psi\rangle\not=0$ the input state $\phi$ is not
present anymore in the output
$\psi={\rm P}_\psi(\phi)$. At first sight this seems to indicate
that through projectors on one-dimensional subspaces there cannot be a
flow of information cfr.~the `wall' in the picture below.
 
\vspace{1.9mm}\noindent
\begin{minipage}[b]{1\linewidth}   
\centering{\epsfig{figure=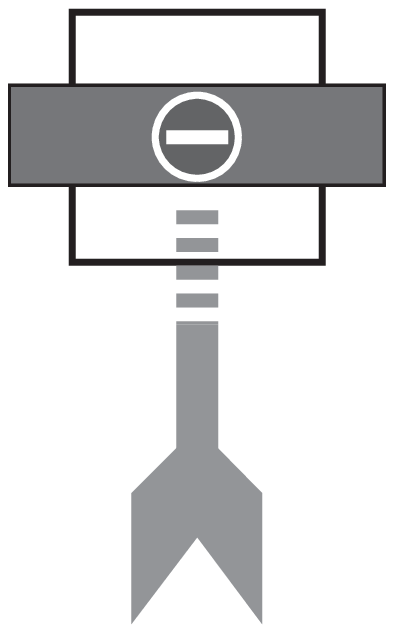,width=27pt}}  
\end{minipage}

\vspace{-0.1mm}\noindent
Theorem \ref{compositionalitytheorem} 
provides a way around this obstacle. 

\vspace{1.9mm}\noindent 
\begin{minipage}[b]{1\linewidth}   
\centering{\epsfig{figure=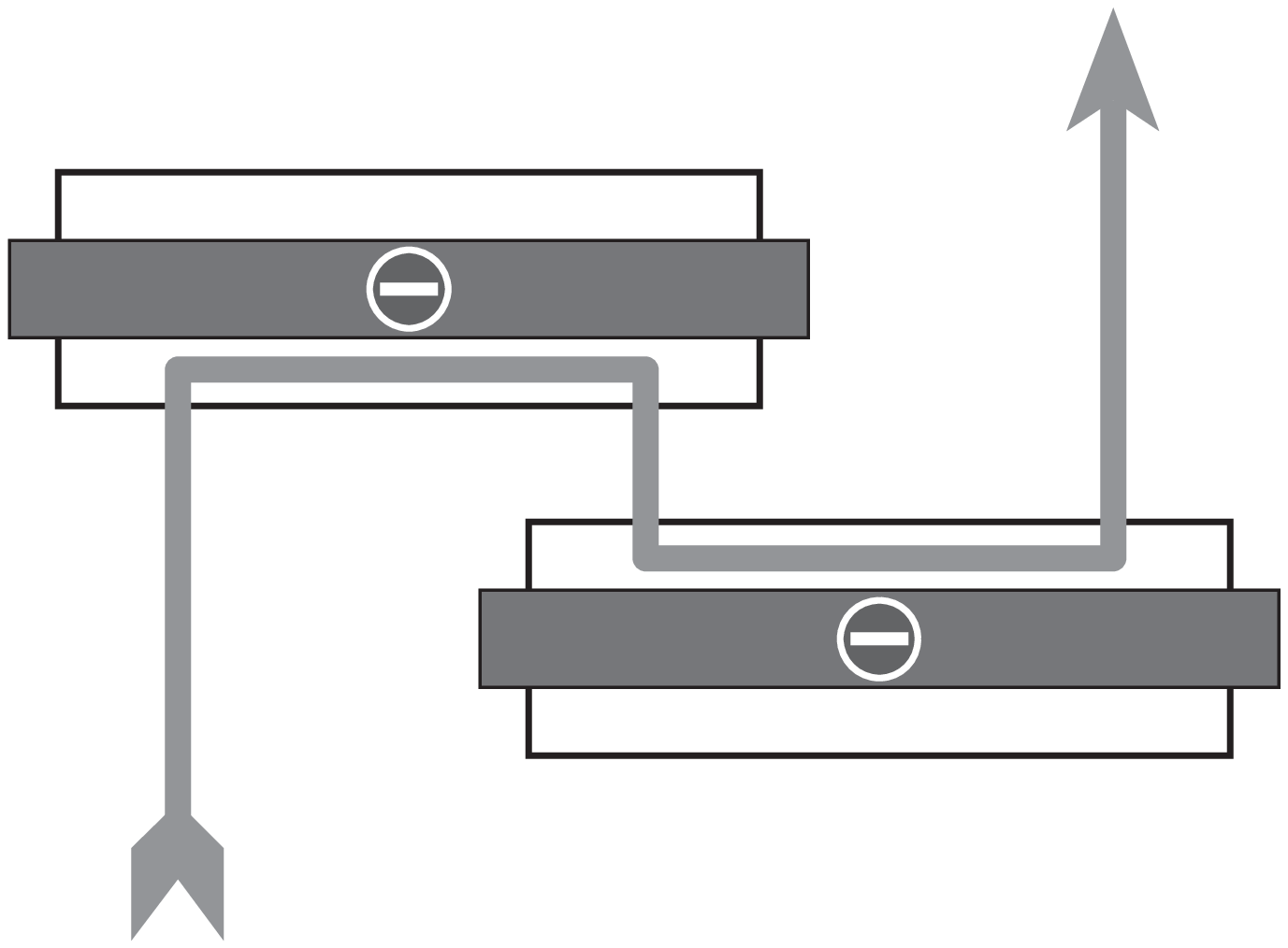,width=95pt}}  
\end{minipage}
 
\vspace{-0.1mm}\noindent
While there cannot be a flow from the
input to the output, there is a `virtual flow'  between the two
inputs and the two outputs of a bipartite projector whenever it is
configured within an appropriate context. And such a bipartite
projector on a state in ${\cal H}_1\otimes{\cal H}_2$ can act on this flow as any
(anti-)linear function $f$ with domain in ${\cal H}_1$ and codomain
in ${\cal H}_2$ --- which is definitely more general than unitary operations
and also more general than actions by (completely) {\it positive
maps}. This behavioral interpretation extends to {\it multipartite
entanglement}, and, as is shown in \cite{RR} \S 6.6, it also enables
interpretation of {\it non-local unitary operations}.

\smallskip
The wall within a projector incarnates the fact that 
\[
{\rm P}_\psi\,\,\stackrel{\rm L}{\simeq}\,\,\psi\otimes\psi\,. 
\]
Indeed, one verifies that {\it disentangled states\,} $\psi\otimes\phi$ are in bijective correspondence 
with those linear maps which have a one-dimensional range \cite{RR}~\S 5.3, that is, since states correspond to
one-dimensional subspaces, disentangled states correspond to (partial) constant maps on states.  Since constant maps
incarnate the absence of information flow (cfr.~`the wall' mentioned above):
\[
{{\rm entangled}\over{\rm disentangled}}
\ \simeq\ 
{{\rm information\ flow}\over{\rm no\ information\ flow}}\,. 
\]
Pursuing this line of thought of conceiving entanglement in terms of its {\it information flow
capabilities\,} yields
a proposal for {\it measuring pure multipartite entanglement\,} \cite{RR}
\S 7.5 --- given a measure for pure bipartite entanglement
e.g.~majorization \cite{Nielsen}.

\smallskip
The use of Theorem \ref{compositionalitytheorem} in Sections
\ref{sec:teleport} and \ref{sec:outputonly} hints towards {\it
automated design\,} of general protocols involving entanglement.  We
started with a simple configuration which conditionally incarnates
the protocol we want to implement. Conceiving this conditional
protocol as a pair consisting of (i) a single path `from root to leaf' in a tree, and, (ii) a configuration picture, we
can extend the tree and the configuration picture with unitary corrections in order to
obtain an unconditional protocol. It constitutes an interesting
challenge to produce an explicit {\it algorithm\,} which realizes this given an
appropriate front-end design language. 

\smallskip
Elaborating on the results in \cite{AbrCoe} S.~Abramsky and the author have produced an
{\it axiomatic characterization\,} of the in this paper exposed behavioral properties of quantum
entanglement with respect to information flow. Remarkably, the additive feature of a
vector space which gives rise to the notion of superposition and hence to that of
entanglement itself seems not to be crucial with respect to the quantum-flow! In particular, we obtain a similar 
information-flow as the one enabled by quantum entanglement
when replacing `vector space' by `set', `linear map' by `relation' and `tensor product' by
`cartesian product' \cite{AbrCoe2}. Replacing `linear map' by `function' in stead of `relation' would not enable
such an information flow. This is due to a different {\it categorical status\,} \cite{CompCat,maclane} of the cartesian
product in the category of sets and relations as compared to its status in the category of sets and functions. 
The category of relations does fail to have a full teleportation protocol because it has no four-vector Bell-base
\cite{AbrCoe2}.

Recent proposals for fault-tolerant quantum computers of
which the architecture is manifestly different from the circuit model require a different
mathematical setting for programming them and reasoning about them \cite{Brie}.   We are convinced
that the insights obtained in this paper provide the appropriate tool for doing so.

\section*{Acknowledgments} 

We thank Samson Abramsky,  
Howard Barnum, Sam Braunstein, Ross Duncan, Peter Hines, Radha Jagadeesan, Prakash
Panangaden, Mehrnoosh Sadrzadeh and  Vlatko Vedral for useful discussions and references.

\end{document}